\documentclass[sigconf, nonacm]{acmart}

\usepackage{algorithm}
\makeatletter 
\g@addto@macro\fs@ruled{\def\@fs@pre{{\normalcolor\hrule height.8pt depth0pt}\kern2pt}\def\@fs@mid{\kern2pt{\normalcolor\hrule}\kern2pt}\def\@fs@post{\kern2pt{\normalcolor\hrule}\relax}}
\makeatother
\usepackage{algpseudocode}
\usepackage{amsmath}
 
\usepackage{amsthm}
\usepackage{minted}
\usepackage{graphicx}
\usepackage{subcaption}
\usepackage{xcolor}
\usepackage{csvsimple}
\usepackage{balance}
\usepackage{multicol}
\usepackage{bbm}
\usepackage{enumitem}
\usepackage{listings}
\usepackage{hyperref}
\usepackage{xcolor}
\usepackage[dvipsnames]{xcolor}
\usepackage[svgnames]{xcolor}
\usepackage{tcolorbox}
\usepackage{tabularx}
\usepackage{ragged2e}
\newcolumntype{L}{>{\RaggedRight\arraybackslash}X}
\newcolumntype{C}{>{\Centering\arraybackslash}X}
\newtheorem{prop}{\textbf{Proposition}}

\usepackage{xspace}
\newcommand{\Sys}{\textsc{SeqoFlow}\xspace}

\usepackage{pifont}
\usepackage{balance}

\usepackage{caption}
\algtext*{EndIf}
\algtext*{EndFor}
\algtext*{EndWhile}
\algtext*{EndFunction}

\newcommand{\ra}[1]{%
  \begingroup
  \color{RoyalBlue}%
  #1%
  \par
  \endgroup
}
\newcommand{\rb}[1]{%
  \begingroup
  \color{Red}%
  #1%
  \par
  \endgroup
}
\newcommand{\rc}[1]{%
  \begingroup
  \color{PineGreen}%
  #1%
  \par
  \endgroup
}
\let\ra\relax
\let\rb\relax
\let\rc\relax

\usepackage{amsmath,amssymb,amsthm}
\newtheorem{definition}{Definition} 

\newcommand{\bag}[1]{\{\!\{\,#1\,\}\!\}}

\newcommand{\Rd}{\mathrm{Rd}}
\newcommand{\Md}{\mathrm{Md}}

\usepackage{pvldb}

\renewcommand\vldbdoi{XX.XX/XXX.XX}

\renewcommand\vldbavailabilityurl{https://github.com/Seurgul/SeqoFlow/tree/main} 

\begin{document}

\title{Navigating Runtime Dynamism in Semantic Query Serving}

\author{Hengrui Zhang}
\affiliation{%
  \institution{Tsinghua University}
  \city{Beijing}
  \country{China}
}
\email{zhanghen22@mails.tsinghua.edu.cn}

\author{Xupeng Miao}
\affiliation{%
  \institution{Peking University}
  \city{Beijing}
  \country{China}
}
\email{xupeng.miao@pku.edu.cn}

\begin{abstract}
LLM-powered semantic queries increasingly combine relational processing with multiple LLM-based operators over unstructured data. Efficiently serving such queries is challenging due to runtime dynamism at two coupled levels. Across queries, dynamic arrivals and heterogeneous resource demands continuously change scheduling priorities. Within a query, heterogeneous characteristics across fine-grained input batches cause operator costs and selectivities to vary throughout execution, making a statically optimized plan unreliable. These two forms of dynamism create a circular dependency: inter-query scheduling requires estimates of remaining LLM work, while the remaining work depends on runtime plan choices whose costs become observable only during execution.

We present \Sys, a runtime framework that jointly adapts query plans and schedules concurrent semantic queries. \Sys enumerates equivalence-preserving plan alternatives under operator dependencies and decomposes queries into fine-grained units, enabling safe plan adaptation and query preemption at batch boundaries. To support both decisions without exhaustively profiling candidate plans on the target LLM, \Sys uses lightweight speculative estimation to capture runtime costs. These estimates jointly guide plan adaptation and inter-query scheduling, \rb{closing the loop between query optimization and LLM inference. \textcolor{black}{Experiments across diverse query structures, data distributions, and arrival rates show that \Sys substantially reduces end-to-end latency compared to existing semantic query systems, adaptive query processing techniques, and conventional scheduling policies.}}
\end{abstract}

\maketitle

\vldbtopmatter


\section{Introduction}
\label{sec: intro}

LLM-powered data analytics has emerged as a new paradigm, attracting growing interest from both academia and industry~\cite{10.14778/3749646.3749685,liu2025palimpzest,10.1145/3788853.3803093,SQLreima2:online,Enrichda89:online,Machinel49:online, 10.1145/3595360.3595858}. By incorporating LLM-powered semantic operators into analytical queries, these systems allow users to express ad hoc intents such as semantic filtering, information extraction, entity matching, ranking, and summarization. Modern data platforms are consequently evolving from monolithic query engines into hybrid systems that tightly couple relational query execution with GPU-backed LLM inference.

Efficiently serving concurrent semantic queries, however, is complicated by pervasive \textbf{\emph{runtime dynamism}} at two levels. At the inter-query level, online query arrivals, completions, and heterogeneous LLM demands continuously alter which query should be served next. At the intra-query level, heterogeneous characteristics across fine-grained input batches, together with content-dependent model behavior, cause operator costs and selectivities to vary throughout query execution.

For example, the following semantic query assesses antibiotic usage in infectious diseases among Asian pediatric departments by combining two medical research paper datasets~\cite{doosterlinck-etal-2023-biodex}.
At runtime, the selectivity of natural-language predicates can vary substantially, making downstream cost estimation unreliable. Moreover, several operators consume outputs generated by upstream LLM operators, introducing cost correlations that are difficult to predict and detect. Executing such queries with a fixed run-to-completion plan can therefore lead to suboptimal latency and unnecessarily delay other concurrent queries sharing the same LLM inference resources. 

\vspace{2 pt}

\begin{lstlisting}[
    language=SQL,
    basicstyle=\ttfamily\scriptsize,
    aboveskip=2pt,
    belowskip=2pt,
    lineskip=-1pt,
    xleftmargin=0pt,
    frame=none,
    columns=fullflexible,
    showstringspaces=false
]
SELECT t2.antibio
FROM (SELECT fulltext_1 FROM table_1
      WHERE sem_filter('{affiliations_1} is an Asian institution.')
        AND sem_filter('{abstract_1} comes from Pediatrics.')) t1
sem_join
     (SELECT sem_map('Assess antibiotic use in {drugs_2}') antibio,
             sem_map('List the drugs studied in {abstract_2}') drugs_2
      FROM table_2
      WHERE sem_filter('{abstract_2} studies infectious diseases.')) t2
ON '{drugs_2} is studied in {fulltext_1}';
\end{lstlisting}

\vspace{2 pt}

Crucially, these two forms of runtime dynamism cannot be addressed independently. Inter-query scheduling requires an estimate of each active query's remaining LLM work. Existing query scheduling techniques typically assume that query execution costs can be estimated or provided before execution and optimize scheduling decisions based on such information~\cite{10.1145/3725319}. However, the remaining work of a semantic query depends on the execution plan selected for its unprocessed input batches. Changing the order or placement of semantic operators may substantially alter intermediate cardinalities and the number and token volume of subsequent LLM invocations. Selecting an efficient plan, in turn, requires operator costs and selectivities that are difficult to predict accurately using static statistics and may become observable only during execution. A runtime plan change therefore revises the query's remaining-work estimate and may alter its scheduling priority relative to other active queries. Subsequent execution exposes new runtime observations that may trigger further plan and scheduling decisions. This creates a \emph{circular dependency} among runtime observation, intra-query plan adaptation, remaining-work estimation, and inter-query scheduling.

Existing techniques address individual parts of this cycle but do not close it end to end. 
Research on expensive predicates and UDFs studies cost-aware operator placement, while adaptive query processing revises execution decisions using runtime statistics. Recent LLM-powered data systems provide logical and physical optimizations for LLM-powered operators~\cite{10.14778/3749646.3749685, liu2025palimpzest, zhu2026beyond}. Separately, database and LLM serving systems support query preemption~\cite{10.1145/3725319}, request scheduling~\cite{kwon2023efficient}, and GPU resource management~\cite{10.1145/3725408}. However, \rb{query optimizers generally select a plan for each query in isolation, without considering how that choice changes the query's remaining work relative to other active queries, whereas LLM schedulers treat inference requests or entire queries as scheduling units without reasoning about alternative query plans. Applying these techniques independently leaves the scheduler with stale remaining-work estimates whenever a query plan changes, and drives each query to commit to a plan before its operator costs become observable.}

In this paper, we argue that navigating runtime dynamism in semantic query serving requires coordinating two individually distinct but mutually dependent decisions: \emph{adapting query plans to intra-query data heterogeneity and scheduling concurrent queries under inter-query workload dynamics}. 
This coordination raises two key challenges.

\vspace{2 pt}

\noindent\textbf{C1: Scheduling under Inter-Query Dynamism.}
As semantic queries arrive, complete, or expose updated estimates of their remaining work, their relative scheduling priorities continuously change. Run-to-completion execution can cause severe head-of-line blocking, whereas scheduling individual LLM requests discards the query structure needed to estimate downstream work. The system therefore needs a scheduling unit that supports timely preemption while retaining sufficient execution context. This raises the first question: \textit{How can the system efficiently switch among concurrent semantic queries and schedule them according to dynamically changing estimates of their remaining work?}

\vspace{2 pt}

\noindent\textbf{C2: Plan Adaptation under Intra-Query Dynamism.}
Fine-grained input batches within a query may exhibit heterogeneous characteristics and execution behavior, causing operator costs and selectivities to vary throughout execution. A statically selected plan may therefore not remain efficient for every batch. However, runtime adaptation must account for input batches that have already passed through parts of the plan, as well as attribute dependencies, operator semantics, and blocking boundaries. This raises the second question: \textit{How can the system safely adapt the execution plan for a query's remaining input batches while preserving equivalence and operator dependencies?}

\vspace{2 pt}

To address these challenges, we present \Sys, a unified runtime framework that \emph{jointly} adapts query plans and schedules concurrent semantic queries. \Sys constructs equivalence-preserving plan alternatives under explicit operator dependencies and decomposes query execution into batch-level \textit{Subquery Execution Units} (SEUs). SEUs provide safe boundaries at which \Sys can incorporate runtime statistics, select a compatible plan for the remaining input batches, preempt an active query, and switch to another query without discarding completed work.
\Sys coordinates plan adaptation and inter-query scheduling through lightweight speculative profiling. It profiles a representative subquery using a smaller language model, collects operator-level selectivity and token-processing costs, and propagates these statistics across compatible plan alternatives to estimate their remaining execution costs. The same estimates guide the global scheduler in reprioritizing active queries, thereby closing the runtime optimization--scheduling loop without exhaustively profiling every candidate plan on the target LLM.
We implement \Sys as an end-to-end semantic query serving system integrated with a GPU-based LLM backend.
Experiments across diverse query structures, batch-level data characteristics, operator dependencies, and arrival rates demonstrate that \Sys substantially reduces end-to-end latency compared with existing semantic query systems, adaptive query processing techniques, and conventional scheduling policies.

In summary, this paper makes the following contributions:
\begin{itemize}[leftmargin=20pt]
    \item We propose \Sys, a unified framework to support efficient serving of semantic queries under runtime dynamism.
    \item We present a runtime adaptation mechanism to capture and select better plans while preserving logical equivalence.
    \item We introduce a fine-grained, structure-aware execution abstraction that exposes safe preemption boundaries and enables dynamic scheduling across concurrent semantic queries.
\end{itemize}

\section{Preliminary and Related Work}

\subsection{Optimizing LLM-powered Data Systems}
\label{prel1}
Recent data systems (e.g., ~\cite{10.14778/3749646.3749685, liu2025palimpzest}) increasingly rely on LLMs as the primary workhorse over unstructured data.
These systems provide declarative interfaces that allow users to program using semantic operators or more general LLM pipelines~\cite{shankar2025docetl}.
In practice, a semantic query is executed by applying each semantic operator to its input data through one or more LLM calls.

Previous work has mainly focused on physical optimization of semantic queries, including accelerating individual operators~\cite{kang13approximate, kang2018blazeit, russo2023accelerating, 10.14778/3749646.3749685}, prompt rewriting~\cite{shankar2025docetl}, and extending the Cascades optimizer~\cite{russo2025abacus}.
Regarding logical optimizations, \cite{mang2026horrilacostbasedplacementsemantic, sun2025agenticdata, liu2025palimpzest, 10.1145/3788853.3803093} directly apply rule-based rewrites, such as operator pull-up, while \cite{zhu2026beyond} prompt LLMs to perform logical rewrites guided by human-written heuristics.
Overall, this literature reflects a shift from treating LLM calls as black-box functions toward principled query models and optimizers for LLM-augmented analytics.
Notably, existing approaches optimize individual, shared-nothing workloads in isolation, while production environment faces more limitations from contentions for bounded resources~\cite{vuppalapati2020building, 10.1145/3725319}, including LLM inference service.
A unified abstraction for expressive semantic querying that coordinates query optimization with preemptive, query-level scheduling remains open.

\subsection{General Serving and Scheduling}
Existing LLM serving systems adopt general-purpose scheduling policies such as FCFS~\cite{kwon2023efficient} or MLFQ~\cite{wu2023fast}, which treat each incoming request as a monolithic unit.
Some systems improve scheduling by leveraging techniques like chunked prefill~\cite{agrawal2024taming}, disaggregation~\cite{patel2024splitwise}, preemption~\cite{wu2023fast}, migration~\cite{sun2024llumnix}, and output length prediction~\cite{qiu2024power}. 

While these approaches are effective for homogeneous inference workloads, they do not directly support semantic queries. 
A semantic query is a structured operator graph whose execution cost depends on operator costs, ordering, and intermediate data sizes.
From a scheduling perspective, this introduces two challenges. 
First, policies such as SRTF require reliable estimates of remaining query time, yet semantic query costs are content-dependent and evolve during execution. 
Second, request-level schedulers also ignore query progress and plan structure. 
Effective scheduling must therefore account for both query-level and operator-level costs.

\subsection{Logical Optimizations in Relational DBMS}
\label{sec:prel2}
Beyond the techniques described in \S~\ref{prel1}, logical query optimization has a rich history, particularly for expensive operators and UDFs that invalidate simple optimization heuristics.
Early work~\cite{hellerstein1993predicate,chaudhuri1999optimization,levy1994query} studies the cost-based placement of predicates and joins, showing that rules such as eager predicate evaluation can become suboptimal when operator costs are non-negligible.
Other work~\cite{neumann2005optimal,10.1007/11788911_10} develops dynamic-programming (DP) algorithms that identify optimal partial plans under known cost and selectivity estimates.
These approaches provide promising correctness within their respective settings, but rely on predefined operator semantics and sufficiently stable cost information. 
They therefore do not directly accommodate fluctuating costs, new query language extensions~\cite{shankar2025docetl,10.1145/3677140}, and latent correlations among operators~\cite{li2024gaussml,kastrati2016optimization,yang2022optimizing}.

Another line of work studies adaptive query processing, which revises execution decisions using observations collected at runtime.
For example, Eddies~\cite{10.1145/342009.335420,10.1145/3725408} are adopted in ML-powered data systems. However, Eddies do not provide a general mechanism for enforcing expressive dependencies among operators, and they require sophisticated implementations and support limited semantics (e.g., blocking operators such as aggregation cannot be easily translated into Eddies' scheme).
Meanwhile, query re-optimization~\cite{kabra1998efficient, zhao2023efficient} mainly considers adapting to a new plan after partial execution of individual queries by re-running the optimization algorithms. 
Therefore, it cannot directly address the highly dynamic characteristics in analyzing unstructured data.
Our goal is to combine the logical optimization's equivalence preserving with the runtime responsiveness of adaptive query processing.

\section{Overview}
\label{sec : ov}

\subsection{Workload Formulation}
We consider an online scenario where analytical queries arrive ad hoc at the system frontend. Each query request $q$ can be formulated as a set of logical operators comprising both traditional relational operators (e.g., filters, joins) and LLM-based semantic operators. Upon arrival, the frontend queues these requests and dispatches them to a centralized backend execution engine, which then interacts with both structured and unstructured data.

Established works that integrate ML/LLM inference within relational data systems \cite{10.14778/3749646.3749685,karanasos2019extending,10.1145/3595360.3595858} as well as production standards~\cite{DataType60:online, pysparks48:online} fetch and process data in fixed-size batches (i.e., multiple rows/tuples) to improve hardware utilization. 
Batches naturally discretize continuous query execution into a sequence of measurable states. Formally, we define the execution state at any given moment as $state = (q, \{op_i\}, B_j)$, representing the query served $q$, the active physical operators $\{op_i\}$, and the current data batch $B_j$.
The continuous interplay of these three components creates a dynamic and complex execution landscape. Because the computational cost of semantic operators depends heavily on the semantic content and input distribution of $B_j$, the optimal execution strategy can shift drastically mid-flight. 
From our observations, a query plan applied to $\{op_i\}$ that was optimal for an early batch may become a severe bottleneck for subsequent batches, requiring dynamic plan mutation. Similarly, the execution time of queries $q$ fluctuates as batches are processed, demanding dynamic scheduler adjustments. 

Despite such volatility, current frameworks fail to exploit the batch-level state granularity that co-optimize all the dimensions. 
Instead, they treat queries as monolithic tasks bound to a run-to-completion execution model. 
This motivates the need for a unified framework capable of dynamically estimating the cost of each transient $state$, unifying both mid-flight query optimization and fine-grained batch-level scheduling.

\subsection{System Design}
\begin{figure}[t!]
    \centering
    \includegraphics[width=0.9\linewidth]{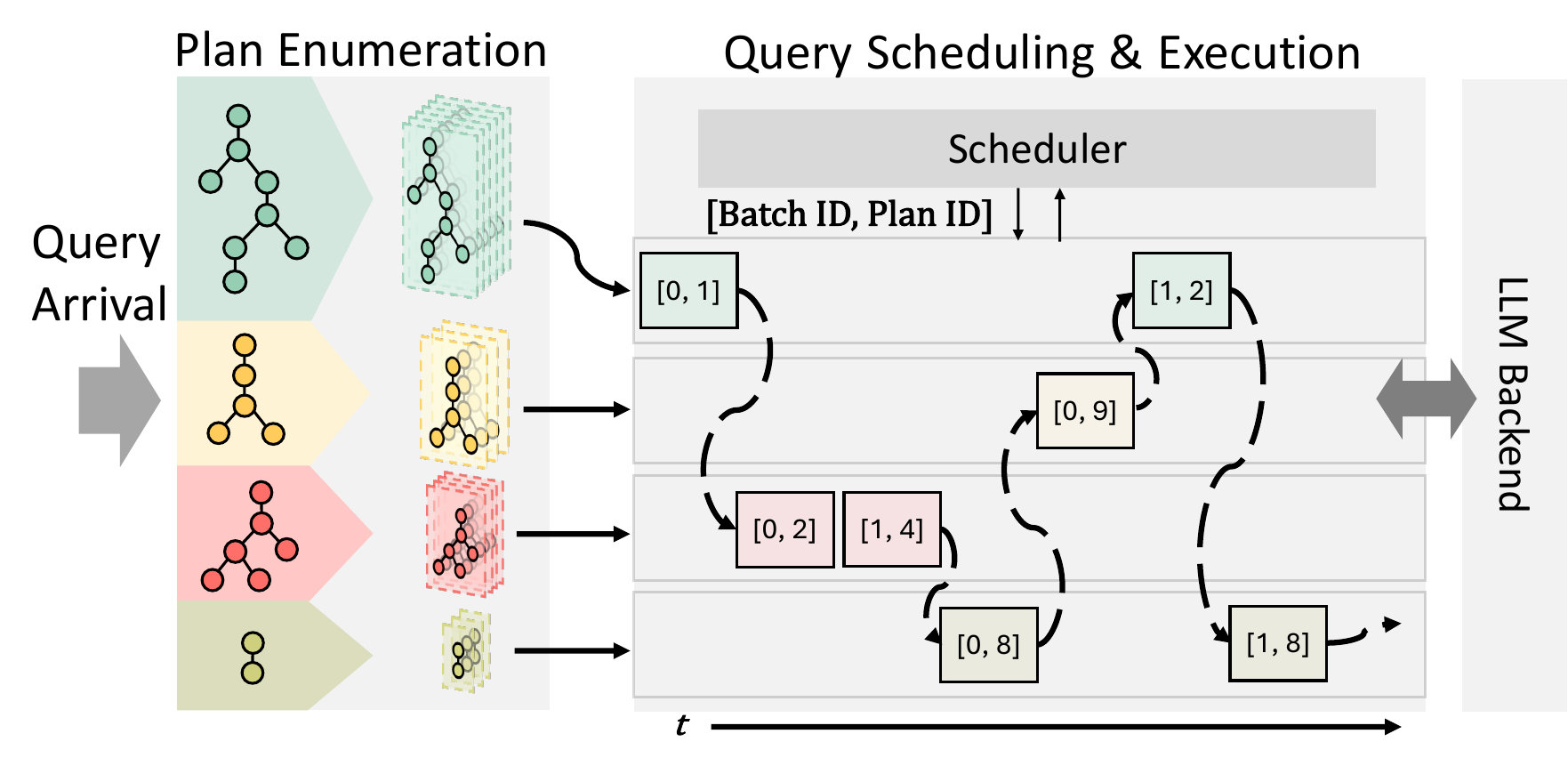}
    \caption{System Architecture -- \textmd{\Sys consists of two stages: A plan enumeration stage that captures equivalent logical plans, and a runtime that schedules and executes batch-level jobs and dynamically selects plan alternatives for batch-level execution.}}
    \label{fig:ov}
\end{figure}
\autoref{fig:ov} provides a high-level overview of \Sys's architecture. The framework consists of two primary components: a static pre-processor that enumerates all equivalent physical plans for a given query $q$ (\S\ref{sec : opt}), and a dynamic execution engine that orchestrates adaptive query processing and inter-query scheduling based on the instantaneous system $state$ (\S\ref{sec: serving}).
Upon query arrival, the pre-process generates and memoizes a comprehensive search space of equivalent query plans. This allows the runtime engine to rapidly retrieve and evaluate alternative execution strategies without re-parsing the query.
During execution, the runtime processes queries in discrete \textit{batch-level preemption units}. This fine-grained architecture enables the global scheduler to safely context-switch and prioritize queries based on dynamic cost metrics, resolving \textbf{C1}.
Simultaneously, to address the intra-query data dynamism in \textbf{C2}, \Sys leverages these units to continuously evaluate the pre-computed search space. By monitoring the current state, \Sys dynamically migrates to the most cost-effective logical plan.

The effectiveness of \Sys's co-design hinges on the ability to accurately estimate costs mid-flight without severely blocking query execution. To bridge this gap, we introduce a novel speculative cost estimation mechanism. By utilizing a lightweight Small Language Model (SLM) to profile transient batch latencies, the system captures real-time costs with high fidelity. 
This speculative feedback loop continuously guides the decisions in scheduling and plan adaptation with bounded computational overhead.


\begin{figure}[t!]
    \centering
    \includegraphics[width=\linewidth]{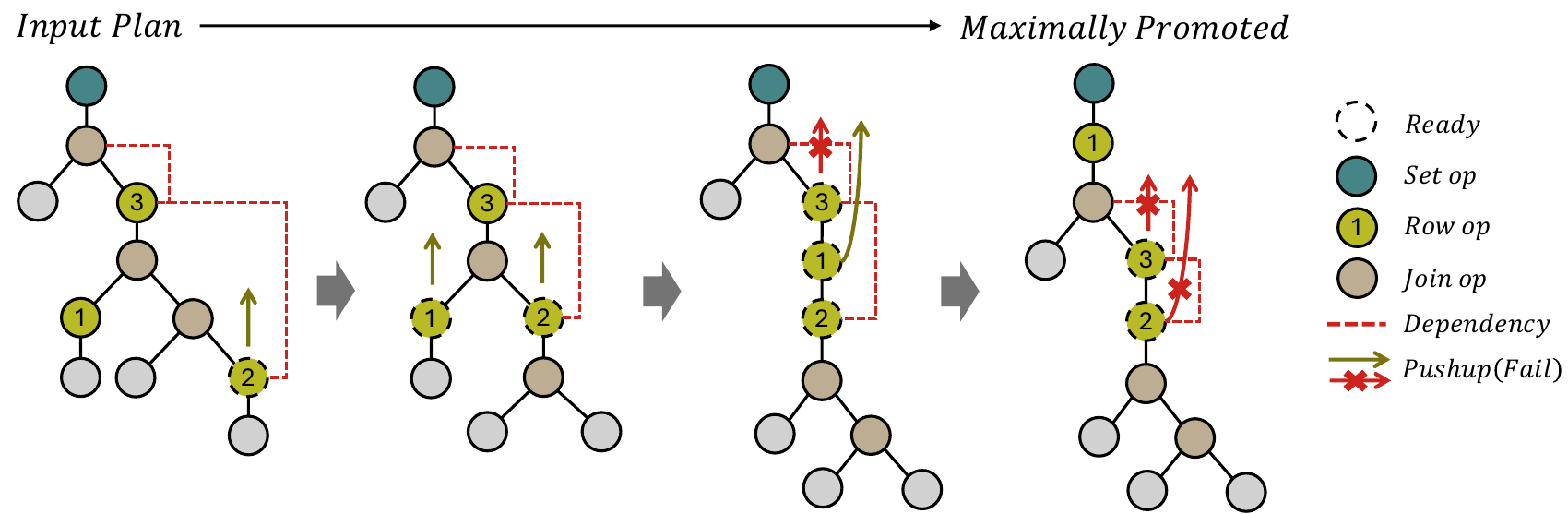}
    \caption{Operator Pushup Phase.}
    \label{fig:pushup}
\end{figure}

\begin{figure}[t!]
    \centering
    \includegraphics[width=0.9\linewidth]{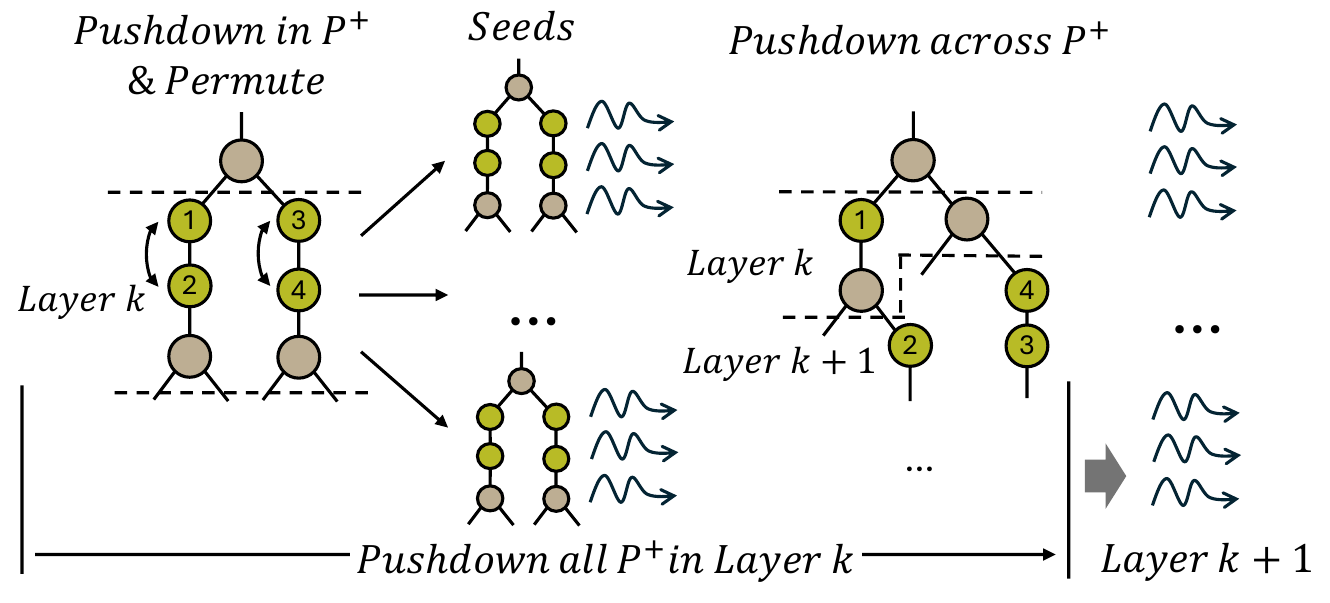}
    \caption{Operator Pushdown Phase.}
    \label{fig:pushdown}
\end{figure}

\section{Intra-Query Plan Enumeration}
\label{sec : opt}


The optimal execution order of semantic operators is not deterministic with simple rule-based heuristics.
Execution performance is largely dependent on the physical cost of LLM-centric procedures and data characteristics, which are not known a priori and can vary substantially.
This introduces significant challenges for transformation schemes applying the predefined set of rules, as they often fail to capture optimal plans falling outside the static patterns.
Unlike prior systems~\cite{mang2026horrilacostbasedplacementsemantic,shankar2025docetl,dai2024uqe,russo2025abacus} that optimize queries by exploring quality–cost trade-offs, we restrict ourselves to lossless optimization and consider only plan transformations that preserve semantic equivalence\footnote{Note that such equivalence is defined with respect to the deterministic semantics of LLM operator intent, ignoring stochastic variation in LLM outputs.}.
In this section, we discuss the algorithms towards an enumeration of equivalent logical plans, enabling the discovery of the best candidate plan structures.

\subsection{Problem Formulation}
\label{sec: prob form}

\subsubsection{Logical Plans}

In prevalent relational query models, a \textit{logical plan} $T = \{op\}$ is a directed tree structure. Each node in the tree represents a logical operator $op$ with at most two children, while edges denote the flow of data between operators.
Leaf nodes correspond to \texttt{scan} operators over base relations (tables), fetching batches of data into physical processing. The root node produces the final output of the query.
Binary operators (\texttt{join} and \texttt{sem\_join}) are the only operators with two children, computing the product of two subtree relation\footnote{While we only discuss binary operators in this paper, the discussion can be extended to any $n$-ary operators without loss of generality.}.
Therefore, they serve as anchors to the plan topology space, which defines the unique tree structure and will not be modified during the transformation process when join reordering is not considered.



\rb{
\begin{definition}
    \label{src}\label{sink}
    For each join $j$, the \textit{source partition} $P^{+}(j)$ consists of $j$ and the operators above it up to the next join or the root. The \textit{sink partition} $P^{-}(j)$ consists of $j$ and the operators below it down to the next joins or scans. A plan without joins forms a single partition.
\end{definition}
}
The source and sink partitions act as the unit of transformations in the divide-and-conquer strategy for plan enumeration in \S\ref{sec:enum}.

\subsubsection{Transformation}
\label{sec : trans}
The optimizer enumerates equivalent plans by iteratively transforming from a seed plan $T$ to a new plan $T'$ through a step-by-step \textit{swap} between parent and child operators, namely $op_{parent}$ and $op_{child}$.
However, to enforce lossless execution, certain swaps are prohibited as they violate the commutativity rules.
Specifically, two rules are considered to determine commutativity:

\vspace{2 pt}
\noindent \textit{\textbf{Rule 1: Data Dependency}} \hspace{2 pt} If $op_{parent}$ consumes a new attribute produced by $op_{child}$, the swap is not allowed as it would raise data dependency error.

\vspace{2 pt}
\noindent \textit{\textbf{Rule 2: Semantic Integrity}} \hspace{2 pt}
Certain operator commutations fundamentally alter intermediate relations, violating the exact semantic equivalence of the query. For example, pushing a selection predicate upstream of an aggregation prematurely filters rows, resulting in an incomplete aggregate computation. To guarantee correctness in plan transformations, we categorize operators into three distinct classes: \textit{row-wise} (e.g., filters, projections), \textit{set-wise} (e.g., aggregations, sorts), and \textit{joins}.
Provided the data dependency constraints of Rule 1 are met, a swap can be applied safely between two consecutive row-wise operators, or a row-wise operator and a join, because these operations evaluate individual rows independently.
To prevent semantic corruption, commutations involving any other class combinations are strictly prohibited.

\ra{
It's noteworthy that the rules are sufficient rather than necessary, as commutations such as pushing a filter on a grouping or sort key also preserve equivalence but lie outside the enumerated space.
Such cases are eliminated to simplify the algorithms' corner cases as it's hardly profitable to inflate a blocking operator's massive gathered input in favor of saving cost from a filter. 
}

\vspace{2 pt}
\noindent \textit{\textbf{Equivalence Guarantee}} \hspace{2 pt}
By strictly following the commutativity rules above, each swap can be considered as an equivalence-preserving operation. \rb{An admissible swap changes the order in which operators apply, and hence intermediate cardinalities and cost, but never the output bag.}
\rb{
Formally, plans $T\equiv T'$ if they produce the same output bag for every valid input.
A row-wise operator $\omega=(p,f)$ applies a predicate $p$ and projection $f$:
$\omega(R)=\bag{f(t)\mid t\in R,\ p(t)}$.
Here $R$ is an input bag. Let $\Rd(\omega)$ be the attributes read by $p$ or $f$, and
$\Md(\omega)$ the attributes created by $f$.

\begin{lemma}[Swap Equivalence]
\label{lem:swap}
(i) Two row-wise operators $\omega_1,\omega_2$ commute if
$\Rd(\omega_i)\cap\Md(\omega_j)=\varnothing$ for $i\ne j$ and
$\Md(\omega_1)\cap\Md(\omega_2)=\varnothing$.
(ii) A row-wise operator $\omega$ can move from above $R\bowtie_\theta S$ to the $R$ branch if
it reads only attributes of $R$ and changes no attribute of $S$ or any attribute used by the join predicate $\theta$. The symmetric condition applies to $S$.
\end{lemma}
\begin{proof}
In case (i), the conditions give $p_i(f_j(t))=p_i(t)$ for $i\ne j$ and
$f_1(f_2(t))=f_2(f_1(t))$ for every row $t$. Both orders therefore retain exactly the same rows satisfying
$p_1(t)\wedge p_2(t)$ and map each to $f_1(f_2(t))=f_2(f_1(t))$.
In case (ii), for rows $r\in R$ and $s\in S$, let $r\circ s$ concatenate $r$ and $s$.
The conditions give $p(r\circ s)=p(r)$,
$f(r\circ s)=f(r)\circ s$, and
$\theta(f(r)\circ s)=\theta(r\circ s)$, which preserve each joined pair and its output row.
\end{proof}

By transitivity of $T \equiv T'$, every plan reachable from the seed plan through a sequence of admissible swaps is equivalent to it, which underpins the correctness of the algorithms in \S\ref{sec:enum}.
The cost of transformations and the size of the plan space are bounded as follows, with proofs in \autoref{app:proofs}.
Let $N$ be the number of operators and $J$ the number of joins. The hook length $h_F(op)$ counts $op$ and its descendants in its constructed Hasse graph $F$.


\begin{prop}[Plan Space]
\label{prop:card}
The number of plans reachable through admissible swaps is at most $\frac{N!}{2^{J}\cdot\prod_{op} h_F(op)}.$
\end{prop}


\begin{prop}[MECE]
\label{prop:cover}
Every plan memoized by Alg.~\ref{alg:pushup} and Alg.~\ref{alg:pushdown} is equivalent to the seed plan and stored once. Without pruning, the memoized plans are \textbf{mutually exclusive and collectively exhaustive} (MECE) over the equivalence-preserving closure defined by rules in \S\ref{sec : trans}. With pruning, for every cost setting, they include a plan of minimum cost.
\end{prop}
}

\subsection{Plan Enumeration}
\label{sec:enum}
The plan enumeration strategy organizes the generation of different plans by first finding an initial state and then spawning equivalent variants from it recursively.
We choose the \textit{maximally promoted tree} as the initial state.
It refers to the topology that all operators cannot be further swapped up with their parents. The process first leverages the \textbf{Operator Pushup} phase to reach this state, {which is unique up to the order of commutative operators within each partition}, then tries spawning equivalent plans by pushing down operators from the initial state, i.e. the \textbf{Operator Pushdown} phase. The procedure is summarized in \autoref{fig:pushup} and \autoref{fig:pushdown}.
\vspace{2 pt}

\noindent \textit{\textbf{Operator Pushup}} \hspace{2 pt}
The goal and intuition of the pushup phase is to promote all row-wise operators to partitions of higher levels until they cannot be promoted anymore. The recursive procedure as present in Algorithm \ref{alg:pushup} and \autoref{fig:pushup} starts by promoting operators in the bottom-most sink partitions to the top.
Each sink partition is a subtree rooted at join operators. The transformation takes each step by selecting a ready operator (denoted \textit{active} in Alg. \ref{alg:pushup}) from the children branches and swapping it up to above the root, or terminating with a failed swap due to commutativity constraints. This is orchestrated inside subroutine \textsc{PushupInSinkPartition}, determining whether all operators in the current partition can no longer be emitted to the parent partition above. If so, the transformed partition can be safely merged into the original plan (line 7 in Alg. \ref{alg:pushup}), and the parent partition starts emitting its operators upward in the next recursion.
Note that pushup mainly operates on partition granularity, and we do not count for the exact order of chained, commutative operators~(e.g., the order of operators 2 and 3 in the final state of \autoref{fig:pushup}) in the same partition.
Different orderings of these chained operators are enumerated in the pushdown phase.
The pushup phase is only used to arrive at a starting point for enumerating all admissible plans later. Therefore, no intermediate state needs to be cached.

\vspace{2 pt}

\begin{algorithm}[t]
\caption{Operator Pushup}\label{alg:pushup}
\begin{algorithmic}[1]

\State \textbf{Input:} Plan $T$

\Function{Pushup}{$P$}
    \State $(P_l, P_r) \gets \Call{GetChildrenSinkPartition}{P}$
    \State \Call{Pushup}{$P_l$}
    \State \Call{Pushup}{$P_r$}

    \State $P' \gets \Call{PushupInSinkPartition}{P}$
    \State $T' \gets \Call{Merge}{T, P'}$
    \State \Return $T'$
\EndFunction

\vspace{6pt}

\Function{PushupInSinkPartition}{$P$}
    \State $P' \gets P$
    \State $active \gets P.sink.children$

    \While{$active \neq \varnothing$}
        \For{$v \in active$}
            \If{\Call{CanSwap}{$v.parent, v$}}
                \State $P' \gets \Call{Swap}{v.parent, v}$
                \State \Call{UpdateDep}{$P'$}
            \EndIf
        \EndFor
        \State $active \gets \Call{GetActiveOps}{ }$
    \EndWhile

    \State \Return $P'$
\EndFunction

\end{algorithmic}
\end{algorithm}

\noindent \textit{\textbf{Operator Pushdown}} \hspace{2 pt}
After the pushup phase, we enumerate the full transitive set under Rules 1 and 2's equivalence guarantee by gradually pushing operators down the maximally promoted state.
A source partition can be considered as a chain of operators with one join at the source (upstream). We assign each partition with its layer in top-down order.
Algorithm \ref{alg:pushdown} and \autoref{fig:pushdown} describes a top-down transformation that pushes operators down to source partitions in lower layers. Operators migrate down (upstream) via swaps, with a memoizer capturing each unique intermediate state along the way.

Intuitively, partitions can spawn variants through two subroutines: \textbf{(1)} Reordering operators within the same partition, and \textbf{(2)} Pushing operators below the source join into lower partitions, namely \textsc{PdInSrcPtt}~(Alg. \ref{alg:pd_in}) and \textsc{PdAcrossSrcPtt}~(Alg. \ref{alg:pd_across}).
The first subroutine permutes chained operators within each parallel layer-$k$ partitions (e.g., in \autoref{fig:pushdown}, there exist two partitions in layer $k$ with two chained operators in each). The Cartesian product of parallel partitions' variants is merged into the plan and memoized. These variants then seed the second subroutine, which records a new variant whenever an operator is pushed below a join. Again, the Cartesian product of these second-stage variants is merged and memoized.
They become the starting plans in the next iteration on layer-$k+1$ partitions.
\ra{
During this stage, \textsc{ApplyPruneHeuristic} (line 26, Alg.~\ref{alg:pd_in}) applies conventional rules, such as predicate pushdown of regular filter, to eliminate dominated candidates. A regular filter costs negligibly compared with an LLM call and never increases cardinality. Any candidate placing it above an LLM operator is therefore dominated by one placing it below, and pruning such candidates never removes the cheapest plan.
}
After iterating over each layer top-down, Alg. \ref{alg:pushdown} is completed with all operators opportunistically pushed-down in the lowest layer and all equivalent plans memoized, except dominated plans removed by pruning. \rb{Without pruning, the enumeration is mutually exclusive and collectively exhaustive (\autoref{app:proofs}).}
\vspace{2pt}

\rb{
\noindent\textit{\textbf{Running Example.}} \hspace{2pt}
For the query presented in \S\ref{sec: intro}, the two filters on \texttt{table\_1} lie in the left sink partition of \texttt{sem\_join}; operators above the join form its source partition.
Pushup promotes eligible row-wise operators toward the join, but the map producing \texttt{drugs\_2} cannot cross it because the join predicate reads that attribute.
Pushdown enumerates both orders of the independent left filters and admissible placements of other row-wise operators.
Every candidate keeps drug extraction before the join and before the map producing \texttt{antibio}.
}

\begin{algorithm}[t]
\caption{Operator Pushdown \label{alg:pushdown}}
\begin{algorithmic}[1]

\State \textbf{Input:} Plan $T$, Memoizer $M$

\Function{Pushdown}{$T, k$}
    \State $\mathcal{P} \gets \Call{GetAllKthLayerPartition}{T, k}$
    \State $seeds \gets \varnothing$

    \For{$P_i \in \mathcal{P}$}
        \State $seeds \gets seeds \cup \{\Call{PdInSrcPtt}{P_i}\}$
    \EndFor

    \State $seeds_2 \gets \varnothing$

    \For{$comb \in \textsc{CartesianProduct}(seeds)$}
        \State $T' \gets \Call{Merge}{T, comb}$
        \State $M \gets M \cup \{T'\}$
        \State $seeds_2 \gets seeds_2 \cup \{comb\}$
    \EndFor

    \State $plans\_next \gets \varnothing$

    \For{$comb \in seeds_2$}
        \For{$comb'' \in \prod_{P'_i \in comb} \Call{PdAcrossSrcPtt}{P'_i}$}
            \State $T'' \gets \Call{Merge}{T, comb''}$
            \State $plans\_next \gets plans\_next \cup \{T''\}$
        \EndFor
    \EndFor

    \For{$T'' \in plans\_next$}
        \State \Call{Pushdown}{$T'', k+1$}
    \EndFor

    \State \Return $M$
\EndFunction

\end{algorithmic}
\end{algorithm}

\begin{algorithm}[t]
\caption{Pushdown in Source Partition \label{alg:pd_in}}
\begin{algorithmic}[1]

\Function{PushdownRowWiseOps} {$P$, $op\_type$}
    \State \Call{UpdateDep}{$P$}
    \State $operators \gets \Call{GetOpsBottomUp} {P.src, op\_type}$
    \For{$op \in operators$}
        \While{\Call{CanSwap} {$op, op.children$} \textbf{and} $op.children$ is not $\bowtie$}
            \State \Call{UpdateDep}{$P$}
            \State $P \gets \Call{Swap} {$op, op.children$}$
        \EndWhile
    \EndFor
    \State \Return $P$
\EndFunction

\vspace{2pt}

\Function{FuseDependencyGroup} {$P$}
    \State $dep\_grps \gets \varnothing$
    \State $op\_ptr \gets P.src.parent$
    \State $anchor \gets op\_ptr$
    \State $current\_grp \gets \{op\_ptr\}$
    \While{$op\_ptr \notin \{\bowtie, \text{stateful ops}\}$}
        \If{\Call{CanSwap}{$op\_ptr, anchor$} \textbf{or} $op\_ptr = anchor$}
            \State $current\_grp \gets current\_grp \cup \{op\_ptr\}$
        \Else
            \State $dep\_grps \gets dep\_grps \cup \{current\_grp\}$
            \State $anchor \gets op\_ptr$
            \State $current\_grp \gets \{op\_ptr\}$
        \EndIf
        \State $op\_ptr \gets op\_ptr.parent$
    \EndWhile
    \State \Return $dep\_grps$
\EndFunction

\vspace{2pt}

\Function{PdInSrcPtt} {Partition $P$}
    \State $P \gets \Call{PushdownRowWiseOps} {P, \{\sigma, \pi, \dots\}}$
    \State $dep\_grps \gets \Call{FuseDependencyGroup} {P}$
    \State $dep\_grps \gets \Call{ApplyPruneHeuristic} {dep\_grps}$
    \State $seeds \gets \cup \Call{ApplyPermute}{dep\_grps}$
    \State \Return $seeds$
\EndFunction

\end{algorithmic}
\end{algorithm}

\begin{algorithm}[t]
\caption{Pushdown across Source Partitions \label{alg:pd_across}}
\begin{algorithmic}[1]

\Function{PdAcrossSrcPtt}{Partition $P$}
    \State $op\_ptr \gets P.src.parent$
    \State $seeds \gets \{P\}$

    \While{$op\_ptr \notin \{\bowtie, \text{stateful ops}\}$}
        \If{\Call{CanSwap}{$op\_ptr, P.src$}}
            \State $P \gets \Call{Swap}{op\_ptr, P.src}$
            \State $seeds \gets seeds \cup \{P\}$
            \State $op\_ptr \gets P.src.parent$
        \Else
            \State $seeds \gets seeds \cup \{P\}$
            \State \textbf{break}
        \EndIf
    \EndWhile

    \State \Return $seeds$
\EndFunction

\end{algorithmic}
\end{algorithm}

\section{Inter-Query Serving}

\label{sec: serving}

\begin{figure*}[t!]
    \centering
    \includegraphics[width=0.85\linewidth]{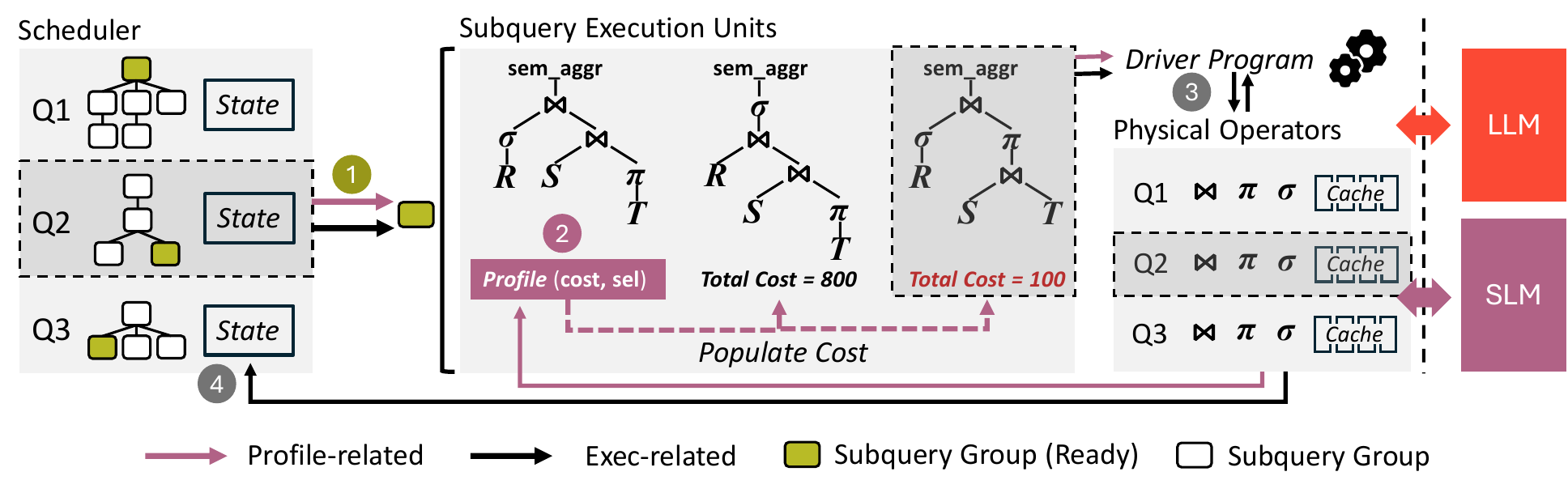}
    \caption{One Complete Iteration of SpecAQP -- \textmd{The workflow consists of 4 stages: 1. The Scheduler picks the next query to be executed and its current ready subquery group; 2. SEU profiles the subqueries to get costs; 3. SEU executes the cheapest subquery in the group; 4. Scheduler updates its internal states for the next iteration.}}
    \label{fig:saqp}
\end{figure*}

\begin{algorithm}[t]
\caption{SpecAQP}\label{alg:saqp}
\begin{algorithmic}[1]

\State \textbf{Input:} Scheduler $\mathit{scheduler}$, Profiler $\mathit{profiler}$

\Function{InitQuery}{$q$}
    \State $q.subq\_grp\_tree \gets \Call{GroupSubquery}{q.candidate\_plans}$
    \State $physical\_ops \gets \Call{InitAllPhysicalOps}{q}$
    \State \Call{Add}{scheduler.queries, q}
    \State \Return $physical\_ops$
\EndFunction

\vspace{6pt}

\Function{Execute}{$ $}
    \While{$scheduler.queries \neq \varnothing$}
        \State $cur\_q \gets \Call{Step}{scheduler}$
        \State $cur\_subq\_grp \gets \Call{GetReadySubq}{cur\_q.subq\_grp\_tree}$
        \State $repr\_subq \gets \Call{GetProfileSubq}{cur\_subq\_grp}$
        \State $driver \gets \Call{GetDriver}{repr\_subq}$

        \State $profile \gets \Call{Run}{driver, repr\_subq, physical\_ops}$

        \For{$subq \in cur\_subq\_grp.subqueries$}
            \State \Call{PopulateProfile}{subq, profile}
        \EndFor

        \State $best\_subq \gets \Call{GetBestSubq}{cur\_subq\_grp}$
        \State $driver \gets \Call{GetDriver}{best\_subq}$

        \State $result, stats \gets \Call{Run}{driver, best\_subq, physical\_ops}$
        \State \Call{Update}{scheduler, result, stats}
    \EndWhile
\EndFunction

\end{algorithmic}
\end{algorithm}

This section presents the serving infrastructure supporting online arrivals of semantic queries. 
\rb{
The middleware exposes a backend-independent execution interface, allowing it to integrate with different inference engines and storage systems.
Backend independence concerns the interface rather than the absence of execution feedback.
The feedback is obtained at batch boundaries (i.e., execution costs) without relying on engine-specific internals.}
As discussed in \S\ref{sec: intro}, the effectiveness hinges on overcoming both \textbf{C1} and \textbf{C2}.
First, it requires support for low-level preemption of in-flight queries, enabling an external scheduler to respond to \textit{inter-query dynamism}.
Second, it can adapt to \textit{intra-query data dynamism} within individual queries by choosing the best candidate plan with respect to varying data input. 
In \S\ref{sec: saqp}, we first introduce the problem of \textbf{Speculative Adaptive Query Processing (SpecAQP)} and the system architecture that resolves both challenges with a unified abstraction of \textbf{Subquery Execution Unit (SEU)}.
In \S\ref{sec: seu}, we discuss the details in SEU, the lowest-level preemptible units in the system. 
Finally, we demonstrate the framework enabling a Shortest Remaining Time First (SRTF) policy in \S\ref{sec: sched}.

\subsection{Speculative Inter-Query AQP}
\label{sec: saqp}
While recent works~\cite{10.1145/3595360.3595858,10.14778/3749646.3749685,liu2025palimpzest} actively attempt to adopt on-GPU ML/LLM operators into query processing, they largely overlook system-level integration with the model-serving backend. The gap makes performances highly dependent on hardware-scaling choices and therefore impractical to real-world scenarios.
Aero~\cite{10.1145/3725408} studies adaptive solutions on analytical queries over video databases with respect to GPU-side optimizations, but still assumes uninterrupted execution of individual queries with exclusive GPU occupancy.

Handling concurrent arrival and execution of queries is non-trivial mainly due to two complexities. First, the orchestration inside individual query's execution is often sophisticated (e.g., ~\cite{10.1145/3725408} applies a specialized implementation of Eddies~\cite{10.1145/342009.335420} to only resolve filter reordering).
Second, the presence of stateful operators (e.g., aggregation requires access to all possible data before producing any output) blocks streaming execution and requires additional state-preserving and recovering.

\vspace{2 pt}
\noindent \textbf{\textit{Subquery Groups}} \hspace{2 pt} 
SpecAQP tackles both challenges via disaggregated execution over \textit{subqueries}. Subqueries are defined as logical subtrees rooted at a single blocking operator with all descendant nodes being non-blocking (e.g., filters, joins). A subquery is the maximal unit that allows non-blocking operators to stream batches to the root and safely yield execution upon preemption.
The root operator can preserve its state simply by buffering the incoming batches upon preemption, incurring no extra memory overhead. Once all batches are buffered, the root evaluates the blocking operation and materializes the result for downstream use.

Subqueries are preemptible at batch-level granularity. Consequently, the original logical plan can be decomposed into an $n$-ary tree hierarchy (i.e. \textit{subquery trees}) of these preemptible subqueries by partitioning on blocking operators.
As discussed in \S\ref{sec: prob form}, blocking operators are strictly non-commutative, serving as rigid boundaries against the relocation of other operators. Because of this structural constraint, the equivalent plans enumerated in \S\ref{sec:enum} share an isomorphic subquery tree topology. Consequently, each topological slot maps to a localized space of equivalent subquery variants from the plans, i.e. \textit{subquery groups}. The subqueries in a group share the same set of local operators.

\vspace{2 pt}
\noindent \textbf{\textit{Subquery Execution Unit (SEU)}} \hspace{2 pt}
Subquery group seamlessly supports SpecAQP's disaggregated execution. 
It enables intra-query AQP where a cost model evaluates subquery variants per batch and picks the best one to execute. 
It also supports inter-query dynamism, allowing execution to be preempted and later resumed using any equivalent subquery in the group. 
To manage this dual-level flexibility, SpecAQP assigns each group an SEU, which we detail in \S\ref{sec: seu}.

\vspace{2 pt}
\noindent \textbf{\textit{Workflow}} \hspace{2 pt}
\autoref{fig:saqp} illustrates the architecture of SpecAQP in a complete scheduler iteration during execution. 
The scheduler is responsible of keeping track of all query requests from users and deciding the query to be executed in the next iteration. 
The whole workflow consists of the following four steps.

\textbf{(1) Select the ready subquery group.} 
After the scheduler has selected the current query through its scheduling algorithm, we identify the subquery group in it that is \textit{ready} to be executed.
This is computed by traversing the $n$-ary tree of groups in depth-first order to find the rightmost subquery group that is not completed yet with all its children groups marked as completed.
Once the \textit{ready} group is found, we create an SEU for it.

\textbf{(2) SEU profiles subquery group with SLM.}
Before an SEU starts execution of a subquery group, it needs to determine which subquery variant in it is the cheapest without causing serious execution blocking. 
Inspired by prior works on speculative inference~\cite{leviathan2023fast,miao2024specinfer} and proxy approximation~\cite{kang13approximate,kang2021accelerating}, we run a \textit{representative subquery} for the next batch using the physical operators on a speculative small language model (SLM).
The SLM is efficient and incurs minimal cost compared to the LLM used for execution, but it offers an estimate for per-operator cost and statistics (e.g., token counts, latency, fan-in/out).
SEU then populates all subqueries in the group, computing their costs using the operator profiles.

\textbf{(3) SEU executes subquery group with LLM.}
Once all subqueries' costs can be determined, the SEU selects the cheapest one to be executed on the LLM, which fetches exactly one batch, processes it and suspends. If this is the last batch, the SEU marks the current group as completed and evaluates the root operator to yield the final result.
It then creates a mock scan operator upon the materialized result, and inserts it into the downstream subquery group as the source of input.

\textbf{(4) Update scheduler state.}
After execution over one batch is completed, the information used in the scheduling algorithm is updated with the current execution statistics, and the scheduler makes the decision for the next iteration.

\subsection{Subquery Execution Unit}
\label{sec: seu}

A Subquery Execution Unit~(SEU) manages the execution of a single data batch once the scheduler activates its corresponding subquery group.
The SEU serves as the basic granularity for cost profiling and execution in the adaptive framework. 
Upon instantiation by the external scheduler, the SEU retrieves the subquery group's operator pool. 
A driver program then links these operators into a physical subquery plan, streaming batch inputs from scan operators or upstream subquery yields through the plan in a depth-first order. 
To facilitate both speculative execution and actual processing, the SEU interfaces with both the SLM and the LLM. 
During cost estimation, it employs \textit{Speculative Profiling}, where it streams a single batch through the SLM to materialize the physical costs of each operator.
During actual execution, the SEU processes the batch identically, but routes it through the LLM.

\vspace{2 pt}
\noindent \textit{\textbf{Populating Costs}} \hspace{2 pt} 
While runtime profiling captures highly accurate execution metrics, exhaustively profiling the exponential search space of equivalent operator topology within a subquery group is prohibitively expensive even with the SLM.
To address this, the SEU profiles only a single representative plan topology (\textit{repr\_subq} in line 11 of Alg.\ref{alg:saqp}), and derives the costs of all alternative equivalent plans via a virtual cost evaluator. 
We select the topology with the maximum pushdown of row-level operators (e.g., filters and projections) as the representative plan. 
\ra{
During speculative profiling, each operator evaluates the \textit{full batch} instead of the rows that pass the previous filtering. The SEU records per-row outcomes of filters and joins as bitmaps and faithfully evaluates the bitwise operations to estimate the intermediate cardinalities of each candidate plan. The evaluator then aggregates the corresponding operator costs according to each candidates' topology.
We detail the profiler design in \autoref{app:prof}.
}
\rb{
After doing so, within the SEU,  the topology with lowest cost and execute its physical plan is picked, as described in line 13-16 of Alg.\autoref{alg:saqp}.
Notably, this full-batch profiling can lead to substantially more work as it forgoes reduction of input cardinality from upstream operators. Ideally, using a fast yet performative SLM helps mitigate this profiling overhead.
To further control profiling cost, \Sys supports both periodic profiling and profiling triggered by a large mismatch between estimated and observed execution costs. 
}

\subsection{SRTF Scheduling}
\label{sec: sched}
Shortest Remaining Time First (SRTF) scheduling is a simple yet theoretically optimal algorithm to minimize latency when serving ad hoc arrival requests. It reduces the cumulative queue waiting time of each request by always serving the one with the shortest remaining time to completion first.
In most existing frameworks like LOTUS, query execution is non-preemptive, and once a complex physical query plan is dispatched, it must run to completion. This creates severe head-of-line blocking, where newly arrived and short queries are forced to wait for long-running, active queries to finish.
Our system overcomes this limitation by enabling batch-level preemption mid-flight.
Because the SEU processes data and materializes costs at the granularity of individual batches, the scheduler can possess an accurate and dynamically updated estimate of the remaining execution time for all active subquery groups~(line 18 in Alg.\autoref{alg:saqp}).
This fine-grained, cost-aware context switching allows our system to natively support SRTF.

\section{Implementation}
\label{sec: impl}

We built \Sys with 6K lines of Python code. We implement the semantic operators following \cite{10.14778/3749646.3749685} and adopt Pandas DataFrame~\cite{reback2020pandas} as the base relational data model.
In this section, we discuss several implementation details.

\vspace{2 pt}

\noindent\textbf{Materialized Nested-Loop Join.}
Streaming joins such as the Symmetric Hash Join~\cite{wilschut1991dataflow} and Ripple Join~\cite{haas1999ripple} 
maintain symmetric state for both inputs and require complex eviction policies at scale. To support batch-level yielding, we instead implement a materialized nested-loop join~\cite{graefe1993volcano, selinger1979access}, 
consistent with existing semantic join implementations~\cite{10.14778/3749646.3749685, trummer2025implementing}. 
It caches only the active outer batch while progressively materializing the inner relation, so a preempted join resumes without refetching its inner input.

\vspace{2pt}
\noindent\textbf{Caching, Broadcasting, and Fingerprinting.}
Mid-flight plan changes alter the state buffered inside joins: once an operator moves across a join, the cached inner relation cannot be replayed blindly without redundant or missing semantic processing. Each batch therefore carries lineage metadata maintained by three primitives. \emph{Caching} buffers per-batch results of row-wise operators for reuse. \emph{Broadcasting} evaluates an operator only on rows with distinct values in its input columns (e.g., a filter's predicate key) and copies the results to duplicate rows. \emph{Fingerprinting} records the set of operators each batch has visited, similar to Eddies~\cite{10.1145/342009.335420}.

\vspace{2 pt}
\noindent \textbf{\textit{Scheduler Cost Modeling}} \hspace{2 pt}
The execution cost of each operator is formulated as a weighted sum of its prefill and decode tokens. To support SRTF scheduling, the system maps this cost to remaining time using dynamically profiled scaling factors.
Specifically, at scheduling step $k$ and wall time $t$, SRTF selects ${\text{argmin}_i} \hspace{2pt} \text{score}_i(t)$-th query to be executed in the next step. The score is defined by:
\begin{flalign*}
&r_{i,k} = \xi \cdot \left(\gamma_1 \cdot \text{prefill} + \gamma_2 \cdot \text{decode} + \gamma_3\right), \quad \phi_k = \mathbbm{1}[k>0] && \\
&\hat R_{i,k} = \phi_k\Big[(1-\beta)\, \big( \hat R_{i,k-1} - (s_{i,k}-s_{i,k-1})\big)\Big] + \beta^{\phi_k}\, r_{i,k} && \\
&\text{score}_i(t) = \hat R_i(t) - \alpha\bigl(t - a_i - s_i(t)\bigr) &&
\end{flalign*}

\noindent $r_{i,k}$ estimates token costs. $\hat R_{i,k}$ models actual remaining work by applying a residual smoothing factor $\beta$.
$s_{i,k}$ or $s_i(t)$ is the total attained service of the $i$-th query at $t$ (step $k$), $a_i$ is the arrival time.
To prevent starvation of long-running queries in SRTF, we apply an aging factor $\alpha$ penalizing long waits with deficient attained service. 
After step $k$, only $\xi$ is updated to match newly completed work $\hat R_{i,k-1} - r_{i,k}$ with $s_{i,k}-s_{i,k-1}$, i.e. the newly attained service.
This online calibration allows remaining-time estimates to reflect the backend's observed execution performance, incorporating changes in inference conditions into subsequent scheduling decisions.

\begin{figure*}[t!]
    \centering
    \includegraphics[width=\linewidth]{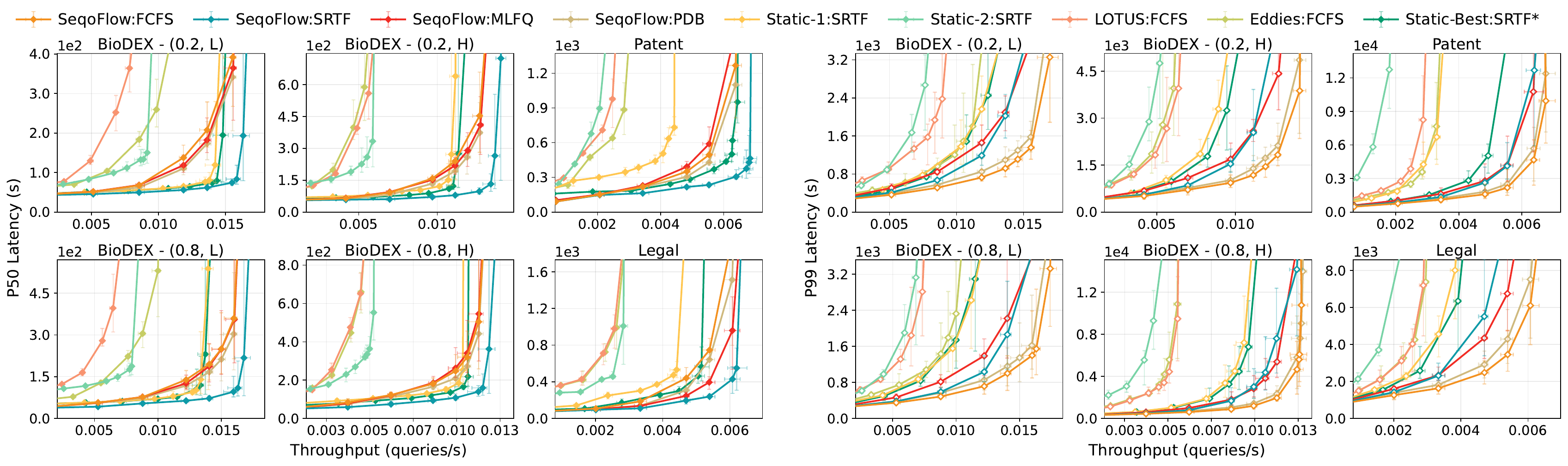}
    \caption{
    \rb{
    Latency vs Throughput in continuous arrivals -- \textmd{P50 and P99 of completed queries after certain cutoff in continuous runs. Runtime settings are denoted as $\{$Method : Scheduler$\}$. Geometric means across 10 runs are reported with min--max span.}
    }
    }

  \label{fig:e2e}
\end{figure*}

\section{Evaluation}
\label{sec:eval}

\subsection{Experiment Setup}
\label{sec: setup}

\noindent \textbf{\textit{Workload Generation}} \hspace{2 pt}
To our knowledge, no existing benchmark stress-tests the runtime dynamism in semantic queries. 
We therefore develop a configurable testbed that generates diverse settings.
We utilize BioDEX \cite{doosterlinck-etal-2023-biodex}, a semantically rich, high-dimensional biomedical dataset to construct controlled, representative and holistic evaluation.
First, 9 representative query templates from the TPC-H~\cite{tpch} benchmark covering a wide range of common patterns in plan topology are selected as base plans, while we replace parts of their regular operators with LLM-based operators.
We then configure these workloads with 2 parameters: the \textbf{variances of selectivities ${\delta(\sigma_{op})}$} and \textbf{sequence lengths ${\delta({seq\_len})}$} among individual operators. Additionally, the selectivity of the entire query $\mathcal{S}$ is used to control the feasible space of the two parameters.




\begin{figure*}[t!]
    \centering
    \includegraphics[width=\linewidth]{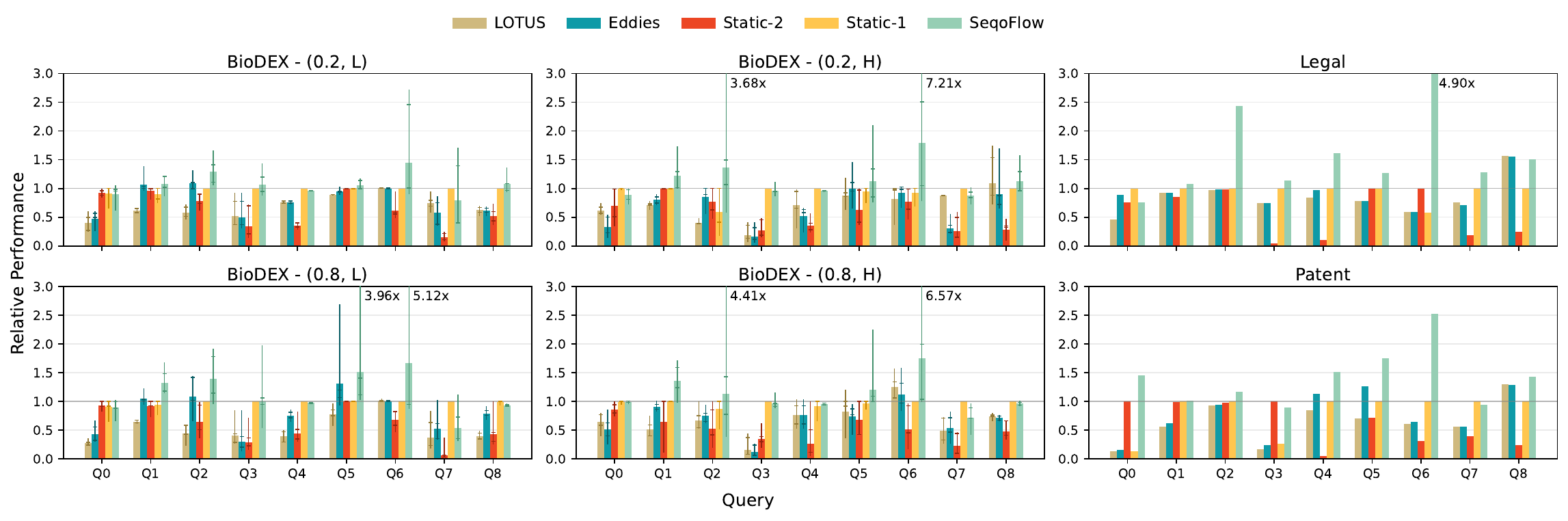}
    \caption{
    \rb{
    Single-query optimization -- \textmd{\Sys can optimize a single query via plan enumeration and adaptively selecting the best plan at runtime. Relative performance is calculated as $min (Static\-1, Static\-2) / latency$, i.e. higher is better. Geometric means across 10 runs are reported with min–max span and 25th/75th percentiles on whiskers.}
    }
    }
    \label{fig:static}
\end{figure*}

\begin{figure}[t!]
    \centering
    \includegraphics[width=0.85\linewidth]{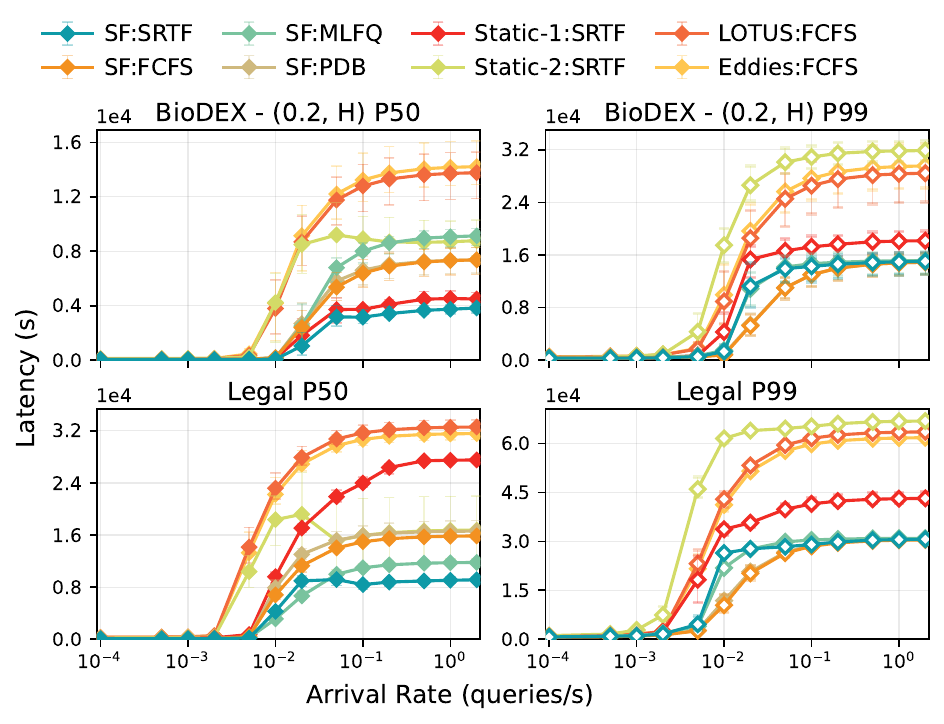}
    \caption{
    \rb{
    Finite arrivals with varying arrival rates -- \textmd{SF stands for \Sys.}
    }
    }
    \label{fig:finite}
\end{figure}

For input (or prefill) sequence lengths, we partition the input space into 3 tiers: \textbf{H}igh, \textbf{M}edium and \textbf{L}ow $\delta(seq\_len)$. A workload is constructed by sampling sequence lengths uniformly, with 50\% of operators drawn from the lower interval and 50\% from the upper interval of the selected tier. 
Then, given ${\delta(\sigma_{op})}$, the selectivities of selective operator (filters and joins) are randomly generated by the testbed to produce a target end-to-end query selectivity $\mathcal{S}$. 
We consider (1) $\delta(\sigma_{op})=0.2$ and (2) $\delta(\sigma_{op})=0.8$, reflecting queries with more or less discrepant selectivity ranges respectively.
In the following sections, unless otherwise noted, controlled settings are denoted as ($\delta(\sigma_{op})$, $\delta(seq\_len)$)\footnote{We present results under High and Low settings due to space constraints. For all experiments in \S\ref{sec:eval} without specifying testbed configurations, we use BioDEX-(0.2, H).}.

\ra{
To test the system against real-world scenarios, we extend the selected query templates to 2 additional corpora with only the definition (i.e., instructions to the LLM) changed within each operator, namely Legal~\cite{butler-2025-open-australian-legal-corpus} and Patent~\cite{allenaiu48:online}. However, in these workloads, no configuration is attached. The cost and selectivity of each operator are determined solely by the actual semantics.
}

\vspace{2 pt}
\noindent \textbf{\textit{Baselines}} \hspace{2 pt}
We compare system performance with the following \textbf{equivalence-preserving} baselines for comparison.

\textbf{LOTUS}~\cite{10.14778/3749646.3749685}.
This baseline features a declarative system with predefined semantic operators. 
To isolate the performance of query execution and scheduling, we disable LOTUS's model cascading feature, ensuring a strictly lossless execution using the identical underlying LLM. For query evaluation, the LOTUS baseline executes the default static logical topology provided by the initial query inputs, as it does not natively support mid-flight plan optimization.

\textbf{Eddies}~\cite{10.1145/342009.335420}.
This approach proposes an Eddy operator, a runtime router that forwards data to different operators according to their respective cost statistics.
It allows adaptively reordering the operators during execution.
We implement Eddies on top of LOTUS with additional support for batch-by-batch processing.
Since the architecture does not consider operator dependencies, we manually bind operators with dependency constraints into fused operators, and leverage existing approaches~\cite{10.14778/3570690.3570695,10.5555/1083592.1083680,10.1145/1066157.1066171,10.1145/3725408} to decide the score of each operator as $cost/(1-selectivity)$ for routing. 

\textbf{Static-1/2}.
We implement these baselines in \Sys by applying rule-based optimizations adopted from~\cite{sun2025agenticdata, mang2026horrilacostbasedplacementsemantic} and human-curated operator-ordering heuristics~\cite{hellerstein1993predicate} to reorder operators.
We use these plans as strong, workload-specific heuristic baselines.
Specifically, Static-1 uses predicate pushdown and Static-2 pushes semantic operators up to the root of the plan.

\rb{
\textbf{Static-Best*}. 
This baseline executes a single fixed plan throughout, namely the best plan in hindsight among all enumerated candidates, or the first plan selected by \Sys when the plan space is too large to execute exhaustively.
}

\rb{
All baseline implementations adopt related techniques mentioned in \S\ref{sec: impl} to ensure a fair comparison.
We use Qwen3-4B and Qwen3-32B~\cite{yang2025qwen3} for speculative profiling and execution, respectively.
We serve them on two separate vLLM~\cite{kwon2023efficient} instances colocated on an 8$\times$A100 NVIDIA GPU server with tensor parallelism (TP=8), and enable chunked prefill~\cite{agrawal2024taming} with $\leq$4096 tokens per chunk.
\Sys issues LLM calls at the query level and makes no assumption about how the LLM is deployed. 

We include profiling overhead throughout the reported results.

\subsection{End-to-end Performance}
\label{sec : e2e}

\subsubsection{Effectiveness of Inter-Query Scheduling}


\autoref{fig:e2e} compares \Sys with run-to-completion baselines under continuous arrivals, using several scheduling policies implemented in \Sys. Randomized queries arrive as a Poisson process with rate $\lambda = \mathit{load}/E[\mathit{latency}]$, where $E[\mathit{latency}]$ is the mean latency of queries run in isolation across all approaches and random seeds.
After a fixed-time warmup, 1,000 queries arrive in a sufficient window as the main cohort. The final completion window then measures the whole cohort to complete while allowing new arrivals. Results therefore exhibit a 1,000-job stream in the middle of a continuous arrival.
\rb{
Since \Sys supports most preemption-based schedulers, we also implement \textbf{PreemptDB (\textit{PDB})}~\cite{10.1145/3725319}'s policy for comparison.
}

\rb{
\textbf{\Sys boosts system throughput at fixed latency budget as well as system capacity.}
Under the same SRTF scheduler, SeqoFlow sustains more load at a given latency and has higher capacity than static plans.
With the scheduler held fixed at SRTF, \Sys sustains up to 1.9$\times$ higher throughput than Static-1 and up to 4.3$\times$ higher than Static-2 at the same P50 latency, and it saturates at a strictly higher load on every workload.
Saturation of system capacity can be solely attributed to per-query execution cost, as the scheduler cannot change the total work the system must perform. 
\Sys also exceeds Static-Best*, which picks the best fixed plan per workload with hindsight. 
\ra{
No single static plan, even one chosen by an oracle, reaches the capacity that \Sys attains by adapting its plan at runtime, and the advantage is largest on Patent and Legal, where operator costs vary most across inputs.
}

\textbf{SRTF contributes to lower P50 latencies at varying loads.}
Within \Sys, SRTF reduces P50 latency by 2.5--4$\times$ compared to FCFS, MLFQ, and PreemptDB near saturation, while matching them at low load.
This isolates the scheduler's contribution. With the plan held fixed, a scheduling policy can only reorder queued work. Its benefit gradually grows with queueing delay as load increases. 
As load approaches system capacity, FCFS, MLFQ, and PDB rise sharply, while SRTF is more resistant in queuing inflation.
This behavior matched the theoretical expectation of SRTF minimizing mean response time, as serving short queries first keeps most queries from waiting even close to saturation.

\textbf{SRTF supported by \Sys surpasses run-to-completion baselines in P50 and P99 latencies.}
Although SRTF is known for inflating tail latency by starving long jobs, \Sys with SRTF sustains 1.9$\times$ more load than LOTUS and Eddies with FCFS before P99 latency inflates. It also stays within 1.4$\times$ of \Sys with FCFS. First, as \Sys reduces per-query service time, it lowers the entire latency distribution, including the tail. Second, \Sys's SRTF ages waiting queries (\S\ref{sec: impl}), which bounds how long a long query can be deferred and limits the queueing delay that starvation would otherwise add to the tail. 
Selecting efficient plans combined with bounded deferral thus outperforms the baselines, even though they use FCFS which prefers reducing tail latency.

\textbf{\Sys speeds up job completions in finite workloads.} 
\autoref{fig:finite} shows latency against offered arrival rate in a simulated run with 200 arrivals of real traces collected from individual queries. Beyond saturation, latency plateaus because each run issues a finite query stream, \Sys with SRTF achieves the lowest P50 latency on both workloads, which is 3.6$\times$ and 3.5$\times$ lower on both datasets than LOTUS with FCFS.  \Sys's P99 plateau is also 2$\times$ below LOTUS. This confirms that even under sustained overload, aging keeps SRTF's tail latency comparable to FCFS.
}

\subsubsection{Single-Query Optimization.}
We further validate \Sys in terms of automatically deriving a plan with near-optimal performance when data characteristics are relatively controlled during runtime.
As illustrated in \autoref{fig:static}, \Sys consistently outperforms LOTUS across all evaluated settings, achieving speedups of up to $7\times$. Crucially, it reliably matches the best performance of the human-curated baselines (Static-1 and Static-2) and even surpasses them in several configurations. This reveals that plan enumeration (\S\ref{sec : opt}) discovers optimization opportunities that rigid static rules miss, while our cost-based selection (\S\ref{sec: seu}) correctly picks a competitive topology. 
Furthermore, for complex plans with highly variable semantic execution costs, \Sys manifests a promising advantage over simple and localized data-routing like Eddies. Eddies also lacks native support for inter-operator data dependencies and restricts reordering to a limited operator subset, making it struggle to optimize such settings.

\begin{figure}[t!]
  \centering
  \begin{subfigure}{0.51\linewidth}
    \includegraphics[width=\linewidth]{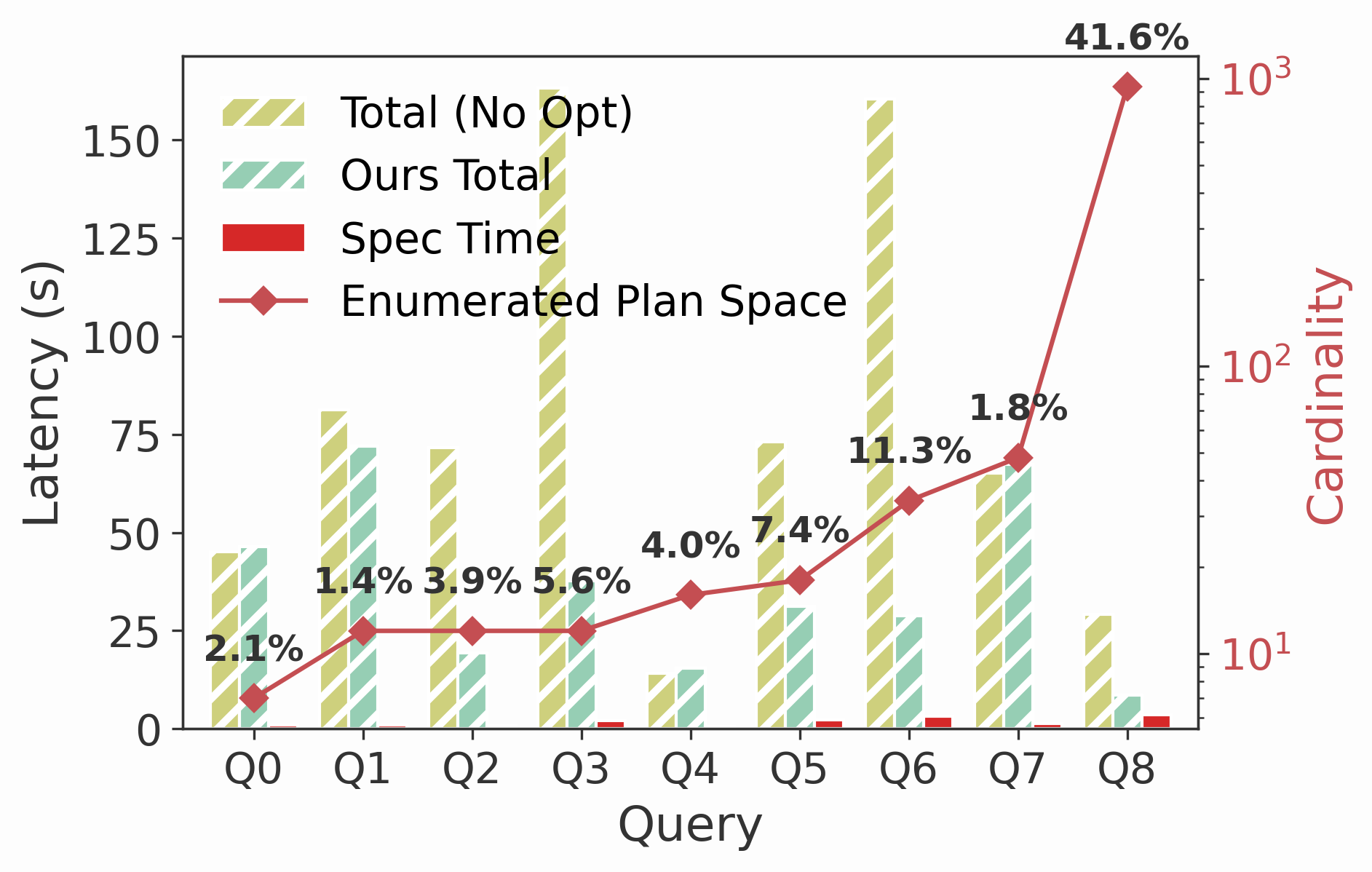}
    \caption{\textmd{Numbers indicate the percentage of E2E system latency attributed to speculative profiling overhead.}}
    \label{fig: lb1}
  \end{subfigure}
  \hfill
  \begin{subfigure}{0.46\linewidth}
    \includegraphics[ width=\linewidth]{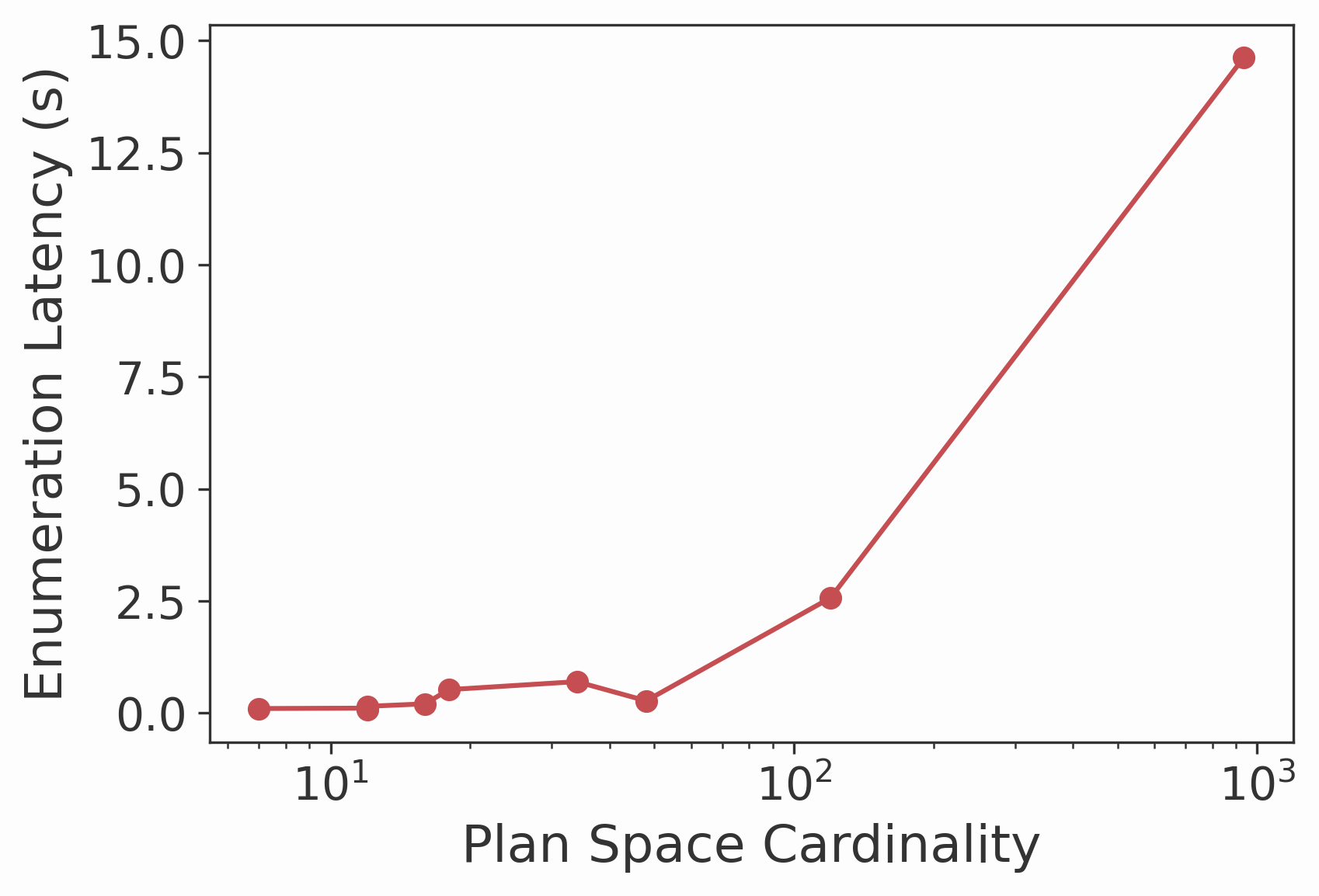}
    \caption{\textmd{The overhead of static plan enumeration with different equivalent plan search space cardinality.}}
    \label{fig: lb2}
  \end{subfigure}
  \caption{Latency breakdown in \Sys}
  \label{fig: lb}
\end{figure}

\begin{figure}[t!]
    \begin{minipage}{\linewidth}
        \centering
        \begin{subfigure}{0.85\linewidth}
            \centering
            \includegraphics[width=\linewidth]{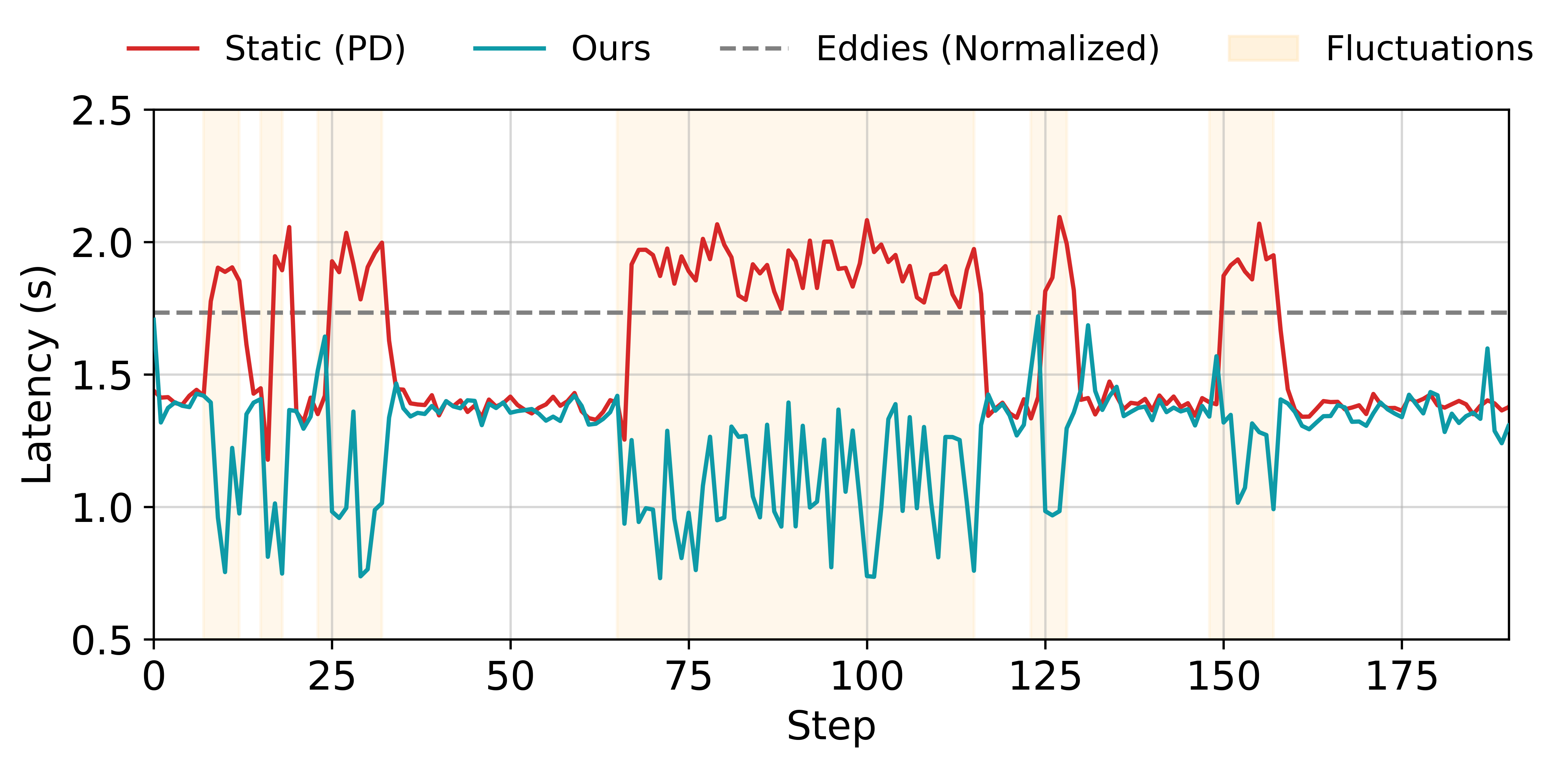}
            \caption{Random continuous fluctuation}
            \label{fig:fluc1}
        \end{subfigure}
        \begin{subfigure}{0.85\linewidth}
            \centering
            \includegraphics[width=\linewidth]{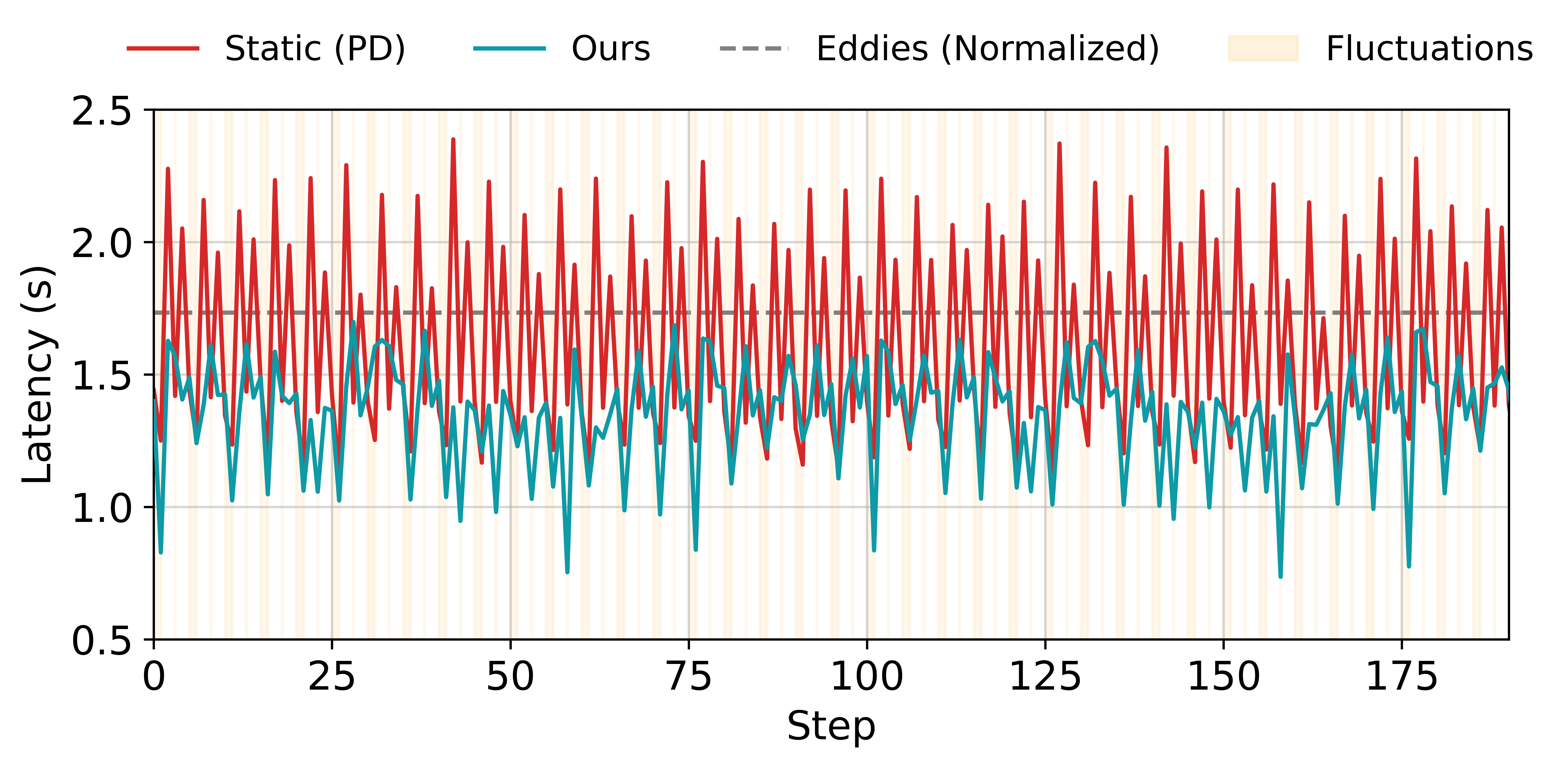}
            \caption{Highly dynamic fluctuation}
            \label{fig:fluc2}
        \end{subfigure}
  \caption{Adaptiveness to fluctuations in workloads -- \textmd{\Sys consistently converges to lower-cost candidate plans when fluctuation happens. Each step processes one batch.}}
  \label{fig:fluc}
  \end{minipage}
\end{figure}

\vspace{2 pt}

\subsubsection{Latency Breakdown}
We further analyze the runtime overhead introduced by \Sys, including plan enumeration and speculative profiling.
As both components depend on the cardinality of the equivalent plan space, we evaluate their overhead under varying search space sizes in \autoref{fig: lb}.
\autoref{fig: lb1} shows that speculative profiling introduces less than 10\% overhead for most queries.
Similarly, \autoref{fig: lb2} demonstrates that plan enumeration contributes negligible compilation overhead in typical cases.
Even for extreme cases with plan spaces containing over 1,000 equivalent plans, the optimization overhead remains justified by the resulting execution time reduction.
For the most challenging case (Q8), speculative profiling accounts for 41.6\% of execution time. However, \Sys still achieves substantial performance improvements compared to executing without optimization.
This efficiency is enabled by \Sys's compact plan enumeration strategy and lightweight profiling mechanism, which avoid exhaustive evaluation of all equivalent plans while preserving accurate cost estimation.

\subsection{Adaptiveness to Intra-Query Dynamism}
\label{sec:adapt}

\begin{figure}[t!]
    \centering
    \includegraphics[width=\linewidth]{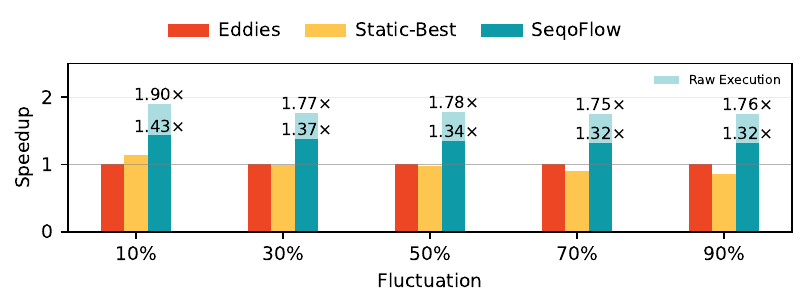}
    \caption{\rb{Speedup over Eddies under fluctuations.}}
    \label{fig:fluc_runtime}
\end{figure}

\rb{
To further test \Sys's adaptiveness to dynamism, the costs and selectivities of selected operators are flipped in a portion~(10\%--90\%) of batches. For instance, a filter that is originally cheap and selective becomes expensive and non-selective, so the plan that is best for the first batch loses its advantage on the flipped batches. \autoref{fig:fluc} demonstrates the per-batch latency of a representative query under both random continuous and highly dynamic fluctuation. \Sys switches to a cheaper plan and returns once the flip in the injected fluctuation windows. 
In \autoref{fig:fluc_runtime}, as the flipped portion grows, Static-Best* steadily degrades and falls below Eddies, since no fixed plan remains optimal once operator costs shift across batches. \Sys re-selects plans batch by batch and outperforms both at every portion, achieving 1.32--1.43$\times$ speedups over Eddies end to end. 
Meanwhile, we show that \Sys can be more effective when only considering execution cost (\textit{Raw Execution)}. This means \Sys's sensitivity to runtime dynamism can be restrained by the SLM when cost drifts are detected frequently, triggering more profiling dispatched to capture the latest costs.
}



\subsection{Validation of Cost Modeling}
\label{sec:cost-valid}

\begin{figure}[t!]
    \centering
    \includegraphics[width=\linewidth]{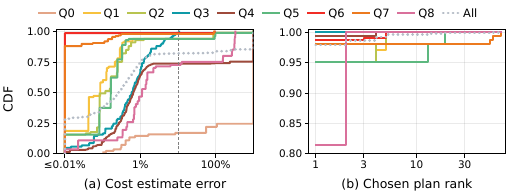}
    \caption{\rb{Accuracy of speculative cost estimation -- \textmd{(a) Relative cost error of each batch, with the dashed line at 10\%. (b) Rank of the chosen plan among all candidates by actual cost.}}}
    \label{fig:error_topk}
\end{figure}

\begin{table}[t]
\centering
\small
\setlength{\tabcolsep}{5pt}
\caption{
\rc{
Ranking quality of estimated against actual remaining time across active queries under 90\% input load.}
}
\label{tab:srtf-quality-90}
\begin{tabular*}{\columnwidth}{@{\extracolsep{\fill}}lcccc@{}}
\toprule
 & \multicolumn{2}{c}{Spearman $\rho$} & \multicolumn{2}{c}{$Acc_{Pair}$} \\
\cmidrule(lr){2-3} \cmidrule(lr){4-5}
SLM/LLM & w/o Age & Age & w/o Age & Age \\
\midrule
4B/32B  & 0.748 & 0.714 & 84.6\% & 74.8\% \\
4B/DeepSeek-Flash  & 0.897 & 0.705 & 93.9\% & 88.2\% \\
4B/gpt-6-luna  & 0.923 & 0.755 & 94.5\% & 94.1\% \\
\bottomrule
\end{tabular*}
\end{table}

\begin{figure}[t!]
    \centering
    \includegraphics[width=\linewidth]{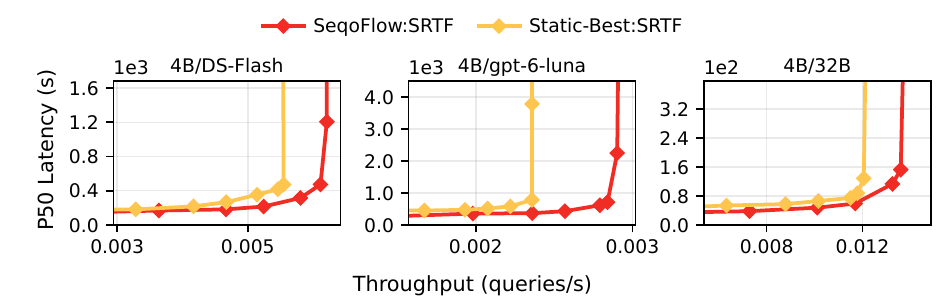}
    \caption{\rb{End-to-end performance across SLM/LLM combinations under SRTF.}}
    \label{fig:different_llms}
\end{figure}

\Sys relies on speculative cost estimates both to select plans within an SEU (\S\ref{sec: seu}) and to prioritize queries under SRTF (\S\ref{sec: sched}). We validate the estimates against each decision.

\noindent\textbf{\textit{Speculative profiling on the SLM reliably selects the best plans for the LLM.}} \hspace{2pt}
Plan selection consumes the estimated cost of each candidate, which compounds operator selectivities with token counts and per-call latency. We therefore measure the relative error between the predicted cost from SLM profiling and the actual cost on the LLM, over all batches of all queries with 70\% of batches flipped (\S\ref{sec:adapt}). Over 80\% of batches are estimated within 10\% error (\autoref{fig:error_topk}a). 
For over 95\% of batches, \Sys executes one of the top-2 plans under actual costs (\autoref{fig:error_topk}b).
}

\rc{
\noindent\textbf{\textit{Speculative estimates rank concurrent queries accurately for SRTF.}} \hspace{2pt}
SRTF depends only on the relative order of remaining times, which \Sys preserves with high fidelity. Without aging, \Sys orders up to 94.5\% of query pairs correctly, with Spearman $\rho$ up to 0.92 (\autoref{tab:srtf-quality-90}). Aging deliberately trades part of this order for bounded waiting, yet keeps at least 74.8\% of pairs correctly ordered.
}

Together, accurate plan selection and query ranking deliver consistent end-to-end gains across models. Pairing the same 4B SLM with official APIs of DeepSeek-Flash~\cite{deepseek2026v41flash}, and GPT-6 Luna, \Sys sustains higher throughput than Static-Best* under every combination, whether the LLM is self-hosted or served through an API (\autoref{fig:different_llms}). Since Static-Best* fixes the best plan in hindsight under the same scheduler, this advantage stems from per-batch plan selection guided by speculative estimates.

\subsection{Case Studies}
\label{sec:cs}
\rc{
\autoref{fig:static} has shown that the benefit of \Sys varies across different workloads. We present a brief study on \Sys's best and worst performances. In the bad cases, Q0 has only 4 operators with few admissible candidates, and it primarily tests whether plan transformations preserve the query result. It therefore leaves little headroom for improvement from discovering the best plan candidate, while speculative profiling can also add overhead.
In contrast, Q1 and Q6 contain multiple semantic filters, projections and joins, yielding more admissible plan candidates with respect to dependency constraints. An early selective filter can avoid downstream projection calls, whereas a costly filter that retains most rows may be better delayed. The favorable order depends on runtime selectivity and token cost, which fixed pushdown or pushup rules cannot determine in advance, making them highly profitable. \Sys uses its runtime estimates to select among these plans, leading to larger gains on these settings.
}

\section{Discussion and Future Work}
\label{sec:lim}

\noindent \textbf{\textit{Concurrency Control.}} \hspace{2 pt}
\Sys naturally supports concurrent OLAP workloads by isolating memory spaces of relevant data structures and cache buffers.
However, extending the system to handle concurrent data modifications and associated consistency challenges remains an interesting direction.

\noindent \textbf{\textit{Search Space Pruning.}} \hspace{2 pt}
\Sys captures equivalent logical plans to adapt to runtime operator behavior.
As with exhaustive plan enumeration, the search space can grow combinatorially. In the extreme case of fully independent operators, the number of equivalent orderings can reach $O(N!)$.
Developing principled pruning techniques that preserve runtime adaptivity while reducing exploration overhead is an important trade-off for future work.
\ra{ With exact costs, exhaustive enumeration guarantees the best plan within the supported space, but finding that plan in practice need not require retaining every candidate.
Besides the well-known pruning heuristics in \S\ref{sec:enum}, advanced runtime pruning could discard plans that are observably poor on current batches and retain promising ones, revisiting them if costs shift.}

\noindent \textbf{\textit{System Integration.}} \hspace{2 pt}
\Sys is designed for storage backends with batch support as discussed in \S\ref{sec : ov}.
It is compatible with existing physical optimizations and orthogonal to the design spaces of LLM serving systems and operator implementations.

\ra{
\noindent \textbf{\textit{Cross-query computation reuse.}} \hspace{2 pt}
Although \Sys improves efficiency in serving multiple semantic queries concurrently, it does not assume any shared operators across different queries as well as the actual LLM calls from them. Exploiting shared work among concurrent queries is previously covered by works such as \cite{bang-2023-gptcache, schroeder2025vcache}, and can be further addressed as a materialized view problem~\cite{hassanzadeh2026semantic} in the multi-query setting.
}




\section{Conclusion}
In this work, we present \Sys, a unified framework that jointly addresses different types of runtime dynamism.
By modeling operator constraints and equivalence boundaries, enabling fine-grained, structure-aware execution, and introducing lightweight speculative cost estimation, our approach effectively adapts to dynamic workloads.
Experimental results show that \Sys significantly outperforms existing methods. 
Overall, this work highlights the importance of co-designing optimization and scheduling for AI-powered data systems and provides a foundation for future semantic query processing engines.


\clearpage
\bibliographystyle{ACM-Reference-Format}
\bibliography{content/ref}

@article{10.1145/3725408,
author = {Kakkar, Gaurav Tarlok and Cao, Jiashen and Sengupta, Aubhro and Arulraj, Joy and Kim, Hyesoon},
title = {Aero: Adaptive Query Processing of ML Queries},
year = {2025},
issue_date = {June 2025},
publisher = {Association for Computing Machinery},
address = {New York, NY, USA},
volume = {3},
number = {3},
url = {https://doi.org/10.1145/3725408},
doi = {10.1145/3725408},
journal = {Proc. ACM Manag. Data},
month = jun,
articleno = {174},
numpages = {27}
}

@inproceedings{10.1145/342009.335420,
author = {Avnur, Ron and Hellerstein, Joseph M.},
title = {Eddies: continuously adaptive query processing},
year = {2000},
isbn = {1581132174},
url = {https://doi.org/10.1145/342009.335420},
doi = {10.1145/342009.335420},
booktitle = {SIGMOD 2000},
pages = {261–272},
numpages = {12}
}

@article{10.14778/3749646.3749685,
author = {Patel, Liana and Jha, Siddharth and Pan, Melissa and Gupta, Harshit and Asawa, Parth and Guestrin, Carlos and Zaharia, Matei},
title = {Semantic Operators and Their Optimization: Enabling LLM-Based Data Processing with Accuracy Guarantees in LOTUS},
year = {2025},
issue_date = {July 2025},
publisher = {VLDB Endowment},
volume = {18},
number = {11},
issn = {2150-8097},
url = {https://doi.org/10.14778/3749646.3749685},
doi = {10.14778/3749646.3749685},
journal = {Proc. VLDB Endow.},
month = jul,
pages = {4171–4184},
numpages = {14}
}

@software{reback2020pandas,
    author       = {The pandas development team},
    title        = {pandas-dev/pandas: Pandas},
    month        = feb,
    year         = 2020,
    publisher    = {Zenodo},
    version      = {latest},
    doi          = {10.5281/zenodo.3509134},
    url          = {https://doi.org/10.5281/zenodo.3509134}
}

@inproceedings{selinger1979access,
  title={Access path selection in a relational database management system},
  author={Selinger, P. Griffiths and Astrahan, Morton M and Chamberlin, Donald D and Lorie, Raymond A and Price, Thomas G},
  booktitle={Proceedings of the 1979 ACM SIGMOD International Conference on Management of Data},
  pages={23--34},
  year={1979}
}

@inproceedings{wilschut1991dataflow,
  title={Dataflow query execution in a parallel main-memory environment},
  author={Wilschut, Annita N and Apers, Peter MG},
  booktitle={Proceedings of the First International Conference on Parallel and Distributed Information Systems (PDIS)},
  pages={68--77},
  year={1991},
  organization={IEEE}
}

@inproceedings{haas1999ripple,
  title={Ripple joins for online aggregation},
  author={Haas, Peter J and Hellerstein, Joseph M},
  booktitle={Proceedings of the 1999 ACM SIGMOD International Conference on Management of Data},
  pages={287--298},
  year={1999}
}

@article{graefe1993volcano,
  title={Volcano - an extensible and parallel query evaluation system},
  author={Graefe, Goetz},
  journal={IEEE Transactions on Knowledge and Data Engineering},
  volume={6},
  number={1},
  pages={120--135},
  year={1994},
  publisher={IEEE}
}

@inproceedings{doosterlinck-etal-2023-biodex,
    title = "{B}io{DEX}: Large-Scale Biomedical Adverse Drug Event Extraction for Real-World Pharmacovigilance",
    author = "D{'}Oosterlinck, Karel  and
      Remy, Fran{\c{c}}ois  and
      Deleu, Johannes  and
      Demeester, Thomas  and
      Develder, Chris  and
      Zaporojets, Klim  and
      Ghodsi, Aneiss  and
      Ellershaw, Simon  and
      Collins, Jack  and
      Potts, Christopher",
    editor = "Bouamor, Houda  and
      Pino, Juan  and
      Bali, Kalika",
    booktitle = "Findings of the Association for Computational Linguistics: EMNLP 2023",
    month = dec,
    year = "2023",
    address = "Singapore",
    publisher = "Association for Computational Linguistics",
    url = "https://aclanthology.org/2023.findings-emnlp.896/",
    doi = "10.18653/v1/2023.findings-emnlp.896",
    pages = "13425--13454"
}

@article{sun2025agenticdata,
  title={Agenticdata: An agentic data analytics system for heterogeneous data},
  author={Sun, Ji and Li, Guoliang and Zhou, Peiyao and Ma, Yihui and Xu, Jingzhe and Li, Yuan},
  journal={arXiv preprint arXiv:2508.05002},
  year={2025}
}

@misc{mang2026horrilacostbasedplacementsemantic,
      title={Horrila: Cost-Based Placement of Semantic Operators in Hybrid Query Plans}, 
      author={Qiuyang Mang and Yufan Xiang and Hangrui Zhou and Runyuan He and Jiaxiang Yu and Hanchen Li and Aditya Parameswaran and Alvin Cheung},
      year={2026},
      eprint={2604.09944},
      archivePrefix={arXiv},
      primaryClass={cs.DB},
      url={https://arxiv.org/abs/2604.09944}, 
}

@article{russo2023accelerating,
  title={Accelerating Aggregation Queries on Unstructured Streams of Data},
  author={Russo, Matthew and Hashimoto, Tatsunori and Kang, Daniel and Sun, Yi and Zaharia, Matei},
  journal={Proceedings of the VLDB Endowment},
  volume={16},
  number={11},
  pages={2897--2910},
  year={2023},
  publisher={VLDB Endowment}
}

@article{kang2018blazeit,
  title={Blazeit: Optimizing declarative aggregation and limit queries for neural network-based video analytics},
  author={Kang, Daniel and Bailis, Peter and Zaharia, Matei},
  journal={arXiv preprint arXiv:1805.01046},
  year={2018}
}

@article{dai2024uqe,
  title={Uqe: A query engine for unstructured databases},
  author={Dai, Hanjun and Wang, Bethany Y and Wan, Xingchen and Dai, Bo and Yang, Sherry and Nova, Azade and Yin, Pengcheng and Phothilimthana, Phitchaya M and Sutton, Charles and Schuurmans, Dale},
  journal={Advances in Neural Information Processing Systems},
  volume={37},
  pages={29807--29838},
  year={2024}
}

@inproceedings{liu2025palimpzest,
  title={Palimpzest: Optimizing ai-powered analytics with declarative query processing},
  author={Liu, Chunwei and Russo, Matthew and Cafarella, Michael and Cao, Lei and Chen, Peter Baile and Chen, Zui and Franklin, Michael and Kraska, Tim and Madden, Samuel and Shahout, Rana and others},
  booktitle={Proceedings of the Conference on Innovative Database Research (CIDR)},
  pages={2},
  year={2025}
}

@article{zhu2026beyond,
  title={Beyond Relational: Semantic-Aware Multi-Modal Analytics with LLM-Native Query Optimization},
  author={Zhu, Junhao and Chen, Lu and Ke, Xiangyu and Fang, Ziquan and Li, Tianyi and Gao, Yunjun and Jensen, Chrisitian S},
  journal={Proceedings of the ACM on Management of Data},
  volume={4},
  number={1 (SIGMOD},
  pages={1--25},
  year={2026},
  publisher={ACM New York, NY, USA}
}

@article{wu2023fast,
  title={Fast distributed inference serving for large language models},
  author={Wu, Bingyang and Zhong, Yinmin and Zhang, Zili and Liu, Shengyu and Liu, Fangyue and Sun, Yuanhang and Huang, Gang and Liu, Xuanzhe and Jin, Xin},
  journal={arXiv preprint arXiv:2305.05920},
  year={2023}
}

@inproceedings{kwon2023efficient,
  title={Efficient memory management for large language model serving with pagedattention},
  author={Kwon, Woosuk and Li, Zhuohan and Zhuang, Siyuan and Sheng, Ying and Zheng, Lianmin and Yu, Cody Hao and Gonzalez, Joseph and Zhang, Hao and Stoica, Ion},
  booktitle={Proceedings of the 29th symposium on operating systems principles},
  pages={611--626},
  year={2023}
}

@inproceedings{qiu2024power,
  title={Power-aware deep learning model serving with $\{$$\mu$-Serve$\}$},
  author={Qiu, Haoran and Mao, Weichao and Patke, Archit and Cui, Shengkun and Jha, Saurabh and Wang, Chen and Franke, Hubertus and Kalbarczyk, Zbigniew and Ba{\c{s}}ar, Tamer and Iyer, Ravishankar K},
  booktitle={2024 USENIX Annual Technical Conference (USENIX ATC 24)},
  pages={75--93},
  year={2024}
}

@inproceedings{sun2024llumnix,
  title={Llumnix: Dynamic scheduling for large language model serving},
  author={Sun, Biao and Huang, Ziming and Zhao, Hanyu and Xiao, Wencong and Zhang, Xinyi and Li, Yong and Lin, Wei},
  booktitle={18th USENIX symposium on operating systems design and implementation (OSDI 24)},
  pages={173--191},
  year={2024}
}

@inproceedings{patel2024splitwise,
  title={Splitwise: Efficient generative llm inference using phase splitting},
  author={Patel, Pratyush and Choukse, Esha and Zhang, Chaojie and Shah, Aashaka and Goiri, {\'I}{\~n}igo and Maleki, Saeed and Bianchini, Ricardo},
  booktitle={2024 ACM/IEEE 51st Annual International Symposium on Computer Architecture (ISCA)},
  pages={118--132},
  year={2024},
  organization={IEEE}
}

@inproceedings{agrawal2024taming,
  title={Taming $\{$Throughput-Latency$\}$ tradeoff in $\{$LLM$\}$ inference with $\{$Sarathi-Serve$\}$},
  author={Agrawal, Amey and Kedia, Nitin and Panwar, Ashish and Mohan, Jayashree and Kwatra, Nipun and Gulavani, Bhargav and Tumanov, Alexey and Ramjee, Ramachandran},
  booktitle={18th USENIX symposium on operating systems design and implementation (OSDI 24)},
  pages={117--134},
  year={2024}
}

@article{russo2025abacus,
  title={Abacus: A Cost-Based Optimizer for Semantic Operator Systems},
  author={Russo, Matthew and Liu, Chunwei and Sudhir, Sivaprasad and Vitagliano, Gerardo and Cafarella, Michael and Kraska, Tim and Madden, Samuel},
  journal={arXiv preprint arXiv:2505.14661},
  year={2025}
}

@article{karanasos2019extending,
  title={Extending relational query processing with ML inference},
  author={Karanasos, Konstantinos and Interlandi, Matteo and Xin, Doris and Psallidas, Fotis and Sen, Rathijit and Park, Kwanghyun and Popivanov, Ivan and Nakandal, Supun and Krishnan, Subru and Weimer, Markus and others},
  journal={arXiv preprint arXiv:1911.00231},
  year={2019}
}

@inproceedings{10.1145/3595360.3595858,
author = {Kakkar, Gaurav Tarlok and Cao, Jiashen and Chunduri, Pramod and Xu, Zhuangdi and Vyalla, Suryatej Reddy and Dintyala, Prashanth and Prabakaran, Anirudh and Bang, Jaeho and Sengupta, Aubhro and Ravichandran, Kaushik and Sivakumar, Ishwarya and Rajoria, Aryan and Raju, Ashmita and Aggarwal, Tushar and Shah, Abdullah and Garg, Sanjana and Suman, Shashank and Kalluraya, Myna Prasanna and Mitra, Subrata and Payani, Ali and Lu, Yao and Ramachandran, Umakishore and Arulraj, Joy},
title = {EVA: An End-to-End Exploratory Video Analytics System},
year = {2023},
isbn = {9798400702044},
publisher = {Association for Computing Machinery},
address = {New York, NY, USA},
url = {https://doi.org/10.1145/3595360.3595858},
doi = {10.1145/3595360.3595858},
booktitle = {Proceedings of the Seventh Workshop on Data Management for End-to-End Machine Learning},
articleno = {8},
numpages = {5},
location = {Seattle, WA, USA},
series = {DEEM '23}
}

@inproceedings{li2024gaussml,
  title={Gaussml: An end-to-end in-database machine learning system},
  author={Li, Guoliang and Sun, Ji and Xu, Lijie and Li, Shifu and Wang, Jiang and Nie, Wen},
  booktitle={2024 IEEE 40th International Conference on Data Engineering (ICDE)},
  pages={5198--5210},
  year={2024},
  organization={IEEE}
}

@misc{tpch,
  title={TPC-H Benchmark},
  author={Transaction Processing Performance Council},
  url={https://www.tpc.org/tpch/},
  year={2026}
}

@inproceedings{hellerstein1993predicate,
  title={Predicate migration: Optimizing queries with expensive predicates},
  author={Hellerstein, Joseph M and Stonebraker, Michael},
  booktitle={SIGMOD 1993},
  pages={267--276},
  year={1993}
}

@article{chaudhuri1999optimization,
  title={Optimization of queries with user-defined predicates},
  author={Chaudhuri, Surajit and Shim, Kyuseok},
  journal={ACM Transactions on Database Systems (TODS)},
  volume={24},
  number={2},
  pages={177--228},
  year={1999},
  publisher={ACM New York, NY, USA}
}

@article{kang13approximate,
  title={Approximate Selection with Guarantees using Proxies},
  author={Kang, Daniel and Gan, Edward and Bailis, Peter and Hashimoto, Tatsunori and Zaharia, Matei},
  journal={Proceedings of the VLDB Endowment},
  volume={13},
  number={11}
}

@article{kang2021accelerating,
  title={Accelerating approximate aggregation queries with expensive predicates},
  author={Kang, Daniel and Guibas, John and Bailis, Peter and Hashimoto, Tatsunori and Sun, Yi and Zaharia, Matei},
  journal={Proceedings of the VLDB Endowment},
  volume={14},
  number={11},
  pages={2341--2354},
  year={2021},
  publisher={VLDB Endowment}
}

@inproceedings{leviathan2023fast,
  title={Fast inference from transformers via speculative decoding},
  author={Leviathan, Yaniv and Kalman, Matan and Matias, Yossi},
  booktitle={International Conference on Machine Learning},
  pages={19274--19286},
  year={2023},
  organization={PMLR}
}

@inproceedings{miao2024specinfer,
  title={Specinfer: Accelerating large language model serving with tree-based speculative inference and verification},
  author={Miao, Xupeng and Oliaro, Gabriele and Zhang, Zhihao and Cheng, Xinhao and Wang, Zeyu and Zhang, Zhengxin and Wong, Rae Ying Yee and Zhu, Alan and Yang, Lijie and Shi, Xiaoxiang and others},
  booktitle={Proceedings of the 29th ACM International Conference on Architectural Support for Programming Languages and Operating Systems, Volume 3},
  pages={932--949},
  year={2024}
}

@article{yang2025qwen3,
  title={Qwen3 technical report},
  author={Yang, An and Li, Anfeng and Yang, Baosong and Zhang, Beichen and Hui, Binyuan and Zheng, Bo and Yu, Bowen and Gao, Chang and Huang, Chengen and Lv, Chenxu and others},
  journal={arXiv preprint arXiv:2505.09388},
  year={2025}
}

@article{shankar2025docetl,
  title={DocETL: Agentic Query Rewriting and Evaluation for Complex Document Processing},
  author={Shankar, Shreya and Chambers, Tristan and Shah, Tarak and Parameswaran, Aditya G and Wu, Eugene},
  journal={Proceedings of the VLDB Endowment},
  volume={18},
  number={9},
  pages={3035--3048},
  year={2025},
  publisher={VLDB Endowment}
}

@article{10.14778/3570690.3570695,
author = {Romero, Francisco and Hauswald, Johann and Partap, Aditi and Kang, Daniel and Zaharia, Matei and Kozyrakis, Christos},
title = {Optimizing Video Analytics with Declarative Model Relationships},
year = {2022},
issue_date = {November 2022},
publisher = {VLDB Endowment},
volume = {16},
number = {3},
issn = {2150-8097},
url = {https://doi.org/10.14778/3570690.3570695},
doi = {10.14778/3570690.3570695},
journal = {Proc. VLDB Endow.},
month = nov,
pages = {447–460},
numpages = {14}
}

@inproceedings{10.5555/1083592.1083680,
author = {Bizarro, Pedro and Babu, Shivnath and DeWitt, David and Widom, Jennifer},
title = {Content-based routing: different plans for different data},
year = {2005},
isbn = {1595931546},
publisher = {VLDB Endowment},
booktitle = {Proceedings of the 31st International Conference on Very Large Data Bases},
pages = {757–768},
numpages = {12},
location = {Trondheim, Norway},
series = {VLDB '05}
}

@inproceedings{10.1145/1066157.1066171,
author = {Babu, Shivnath and Bizarro, Pedro and DeWitt, David},
title = {Proactive re-optimization},
year = {2005},
isbn = {1595930604},
publisher = {Association for Computing Machinery},
address = {New York, NY, USA},
url = {https://doi.org/10.1145/1066157.1066171},
doi = {10.1145/1066157.1066171},
booktitle = {Proceedings of the 2005 ACM SIGMOD International Conference on Management of Data},
pages = {107–118},
numpages = {12},
location = {Baltimore, Maryland},
series = {SIGMOD '05}
}

@article{kastrati2016optimization,
  title={Optimization of conjunctive predicates for main memory column stores},
  author={Kastrati, Fisnik and Moerkotte, Guido},
  journal={Proceedings of the VLDB Endowment},
  volume={9},
  number={12},
  pagess={1125--1136},
  year={2016},
  publisher={VLDB Endowment}
}

@article{trummer2025implementing,
  title={Implementing Semantic Join Operators Efficiently},
  author={Trummer, Immanuel},
  journal={arXiv preprint arXiv:2510.08489},
  year={2025}
}

@inproceedings{levy1994query,
  title={Query optimization by predicate move-around},
  author={Levy, Alon Y and Mumick, Inderpal Singh and Sagiv, Yehoshua},
  booktitle={VLDB},
  volume={94},
  pages={12--15},
  year={1994}
}

@inproceedings{10.1145/3788853.3803093,
author = {Liskowski, Pawe\l{} and Han, Benjamin and Aggarwal, Paritosh and Chen, Bowei and Jiang, Boxin and Jindal, Nitish and Li, Zihan and Lin, Aaron and Schmaus, Kyle and Tayade, Jay and Zhao, Weicheng and Datta, Anupam and Wiegand, Nathan and Tsirogiannis, Dimitris},
title = {Cortex AISQL: A Production SQL Engine for Unstructured Data},
year = {2026},
isbn = {9798400724503},
publisher = {Association for Computing Machinery},
address = {New York, NY, USA},
url = {https://doi.org/10.1145/3788853.3803093},
doi = {10.1145/3788853.3803093},
booktitle = {Companion of the International Conference on Management of Data},
pages = {400–412},
numpages = {13},
location = {India},
series = {SIGMOD Companion '26}
}

@misc{SQLreima2:online,
author = {},
  title = {SQL reimagined for the AI era with BigQuery AI functions},
  url = "https://cloud.google.com/blog/products/data-analytics/sql-reimagined-for-the-ai-era-with-bigquery-ai-functions",
month = {},
year = {},
}

@misc{Enrichda89:online,
author = {},
  title = {Enrich data using AI Functions},
  url = "https://docs.databricks.com/aws/en/large-language-models/ai-functions",
month = {},
year = {},
}

@misc{Machinel49:online,
author = {},
  title = {Machine learning and AI functions in GoogleSQL},
  url = "https://docs.cloud.google.com/spanner/docs/reference/standard-sql/ml-functions",
month = {},
year = {},
}

@article{10.1145/3725319,
author = {Huang, Kaisong and Zhou, Jiatang and Zhao, Zhuoyue and Xie, Dong and Wang, Tianzheng},
title = {Low-Latency Transaction Scheduling via Userspace Interrupts: Why Wait or Yield When You Can Preempt?},
year = {2025},
issue_date = {June 2025},
publisher = {Association for Computing Machinery},
address = {New York, NY, USA},
volume = {3},
number = {3},
url = {https://doi.org/10.1145/3725319},
doi = {10.1145/3725319},
journal = {Proc. ACM Manag. Data},
month = jun,
articleno = {182},
numpages = {25}
}

@inproceedings{vuppalapati2020building,
  title={Building an elastic query engine on disaggregated storage},
  author={Vuppalapati, Midhul and Miron, Justin and Agarwal, Rachit and Truong, Dan and Motivala, Ashish and Cruanes, Thierry},
  booktitle={17th USENIX Symposium on Networked Systems Design and Implementation (NSDI 20)},
  pages={449--462},
  year={2020}
}

@misc{DataType60:online,
author = {},
  title = {Data Types and In-Memory Data Model — Apache Arrow v25.0.0},
  url = "https://arrow.apache.org/docs/python/data.html",
month = {},
year = {}
}

@misc{pysparks48:online,
author = {},
  title = {pyspark.sql.DataFrame},
  url = "https://spark.apache.org/docs/latest/api/python/reference/pyspark.sql/api/pyspark.sql.DataFrame.html",
month = {},
year = {}
}

@inproceedings{neumann2005optimal,
  title={On the optimal ordering of maps and selections under factorization},
  author={Neumann, Thomas and Helmer, Sven and Moerkotte, Guido},
  booktitle={21st International Conference on Data Engineering (ICDE'05)},
  pages={490--501},
  year={2005},
  organization={IEEE}
}

@inproceedings{10.1007/11788911_10,
author = {Neumann, Thomas and Helmer, Sven and Moerkotte, Guido},
title = {On the optimal ordering of maps, selections, and joins under factorization},
year = {2006},
isbn = {3540359699},
publisher = {Springer-Verlag},
address = {Berlin, Heidelberg},
url = {https://doi.org/10.1007/11788911_10},
doi = {10.1007/11788911_10},
booktitle = {Proceedings of the 23rd British National Conference on Databases, Conference on Flexible and Efficient Information Handling},
pages = {115–126},
numpages = {12},
location = {Belfast, Northern Ireland, UK},
series = {BNCOD'06}
}

@article{10.1145/3677140,
author = {Wu, Renzhi and Chunduri, Pramod and Payani, Ali and Chu, Xu and Arulraj, Joy and Rong, Kexin},
title = {SketchQL: Video Moment Querying with a Visual Query Interface},
year = {2024},
issue_date = {September 2024},
publisher = {Association for Computing Machinery},
address = {New York, NY, USA},
volume = {2},
number = {4},
url = {https://doi.org/10.1145/3677140},
doi = {10.1145/3677140},
journal = {Proc. ACM Manag. Data},
month = sep,
articleno = {204},
numpages = {27}
}

@article{yang2022optimizing,
  title={Optimizing machine learning inference queries with correlative proxy models},
  author={Yang, Zhihui and Wang, Zuozhi and Huang, Yicong and Lu, Yao and Li, Chen and Wang, X Sean},
  journal={arXiv preprint arXiv:2201.00309},
  year={2022}
}

@article{zhao2023efficient,
  title={Efficient query re-optimization with judicious subquery selections},
  author={Zhao, Junyi and Zhang, Huanchen and Gao, Yihan},
  journal={Proceedings of the ACM on Management of Data},
  volume={1},
  number={2},
  pages={1--26},
  year={2023},
  publisher={ACM New York, NY, USA}
}

@inproceedings{kabra1998efficient,
  title={Efficient mid-query re-optimization of sub-optimal query execution plans},
  author={Kabra, Navin and DeWitt, David J},
  booktitle={Proceedings of the 1998 ACM SIGMOD international conference on Management of data},
  pages={106--117},
  year={1998}
}

@book{knuth1998art,
  title={The Art of Computer Programming: Sorting and Searching, volume 3},
  author={Knuth, Donald E},
  year={1998},
  publisher={Addison-Wesley Professional}
}

@misc{butler-2025-open-australian-legal-corpus,
    author = {Butler, Umar},
    year = {2025},
    title = {Open Australian Legal Corpus},
    publisher = {Isaacus},
    version = {7.1.0},
    doi = {10.57967/hf/2784},
    url = {https://huggingface.co/datasets/isaacus/open-australian-legal-corpus}
}

@misc{allenaiu48:online,
author = {},
  title = {allenai/us-patents · Datasets at Hugging Face},
  url = "https://huggingface.co/datasets/allenai/us-patents",
month = {},
year = {}
}

@inproceedings{bang-2023-gptcache,
    title = "{GPTC}ache: An Open-Source Semantic Cache for {LLM} Applications Enabling Faster Answers and Cost Savings",
    author = "Bang, Fu",
    editor = "Tan, Liling  and
      Milajevs, Dmitrijs  and
      Chauhan, Geeticka  and
      Gwinnup, Jeremy  and
      Rippeth, Elijah",
    booktitle = "Proceedings of the 3rd Workshop for Natural Language Processing Open Source Software ({NLP}-{OSS} 2023)",
    month = dec,
    year = "2023",
    address = "Singapore",
    publisher = "Association for Computational Linguistics",
    url = "https://aclanthology.org/2023.nlposs-1.24/",
    doi = "10.18653/v1/2023.nlposs-1.24",
    pages = "212--218"
}

@article{schroeder2025vcache,
  title={vcache: Verified semantic prompt caching},
  author={Schroeder, Luis Gaspar and Desai, Aditya and Cuadron, Alejandro and Chu, Kyle and Liu, Shu and Zhao, Mark and Krusche, Stephan and Kemper, Alfons and Zaharia, Matei and Gonzalez, Joseph E},
  journal={arXiv preprint arXiv:2502.03771},
  year={2025}
}

@article{hassanzadeh2026semantic,
  title={Semantic Views: Certified Reuse of LLM Labels Across Semantic Queries},
  author={Hassanzadeh, Oktie and Subramanian, Dharmashankar},
  journal={arXiv preprint arXiv:2610.12768},
  year={2026}
}

@misc{deepseek2026v41flash,
  title={{DeepSeek-V4.1-Flash} Technical Report},
  author={{DeepSeek-AI}},
  year={2026},
  howpublished={\url{https://huggingface.co/deepseek-ai/DeepSeek-V4.1-Flash/blob/main/DeepSeek_V41_Tech_Report.pdf}}
}

\clearpage
\appendix
\section{Proofs of Plan Enumeration Properties}
\label{app:proofs}

Fix a plan $T$ with operator set $V$, $|V|=N$, and $J$ joins.
Operators are labeled; read sets are nonempty and origin-qualified, so
each admissible join crossing selects a unique input branch.
Let $\mathcal R(T)$ be its closure under admissible swaps and
$\mathcal E_0(T)$ its memoized plan set without pruning.

\subsection{Plan Space}
\label{app:suppl-plan-space}

\noindent\textbf{Proposition~\ref{prop:card} (Plan Space).}
For the forest $F$ defined below,
\[
|\mathcal R(T)|\le
\frac{N!}{2^J\prod_{v\in V}h_F(v)}.
\]

\noindent\textbf{Dependency poset.}
Write $u\prec_D v$ for transitive producer--consumer dependence and
$u\prec_Q v$ when $u$ is a proper descendant of $v$ in plan $Q$.
Let $B$ contain the precedence constraints imposed by the fixed skeleton
and noncrossable boundaries. Retaining these constraints from $T$ gives
\[
P=(V,\prec_P),\qquad \prec_P=\operatorname{tc}(\prec_D\cup B),
\]
with $\prec_P\subseteq\prec_Q$ for every $Q\in\mathcal R(T)$,
where $\operatorname{tc}$ denotes transitive closure.
Let $H=\operatorname{Cov}(P)$ be its Hasse edges, directed from smaller to
larger elements. For any poset $A$, write $\mathcal L(A)$ for its linear
extensions and $e(A)=|\mathcal L(A)|$.

\noindent\textbf{Uniform hook construction.}
Let $P=P_1\sqcup\cdots\sqcup P_m$, where each component $P_i$ has a tree
Hasse diagram and a unique maximal or unique minimal element $r_i$.
Dualize precisely the reverse-rooted components:
\[
F_i=\begin{cases}
P_i,&r_i\text{ is maximal},\\
P_i^{{op}},&r_i\text{ is minimal},
\end{cases}
\qquad F=\bigsqcup_{i=1}^m F_i.
\]
For a singleton, either choice agrees. Each $F_i$ is a rooted tree with
unique maximum $r_i$. Its hook is uniformly the subtree size
\[
h_F(v)=|\{u\in P_i:v\text{ lies on the path from }u\text{ to }r_i\}|.
\]
Equivalently, using reflexive orders,
\[
h_F(v)=\begin{cases}
|\{u:u\preceq_{P_i}v\}|,&r_i\text{ is maximal},\\
|\{u:v\preceq_{P_i}u\}|,&r_i\text{ is minimal}.
\end{cases}
\]

\begin{lemma}[Uniform hook formula]
\label{lem:suppl-uniform-hook}
With this construction,
\begin{equation}
\label{eq:suppl-forest-hook}
e(P)=e(F)=\frac{N!}{\prod_{v\in V}h_F(v)}.
\end{equation}
\end{lemma}
\begin{proof}
Given $\lambda\in\mathcal L(P)$, reverse its subsequence on each dualized
component, leaving the positions occupied by that component unchanged.
This involution gives a bijection $\mathcal L(P)\longleftrightarrow
\mathcal L(F)$. Let $n_i=|P_i|$. Applying Knuth's formula~\cite{knuth1998art}
to each rooted $F_i$ and interleaving the components gives
\[
\begin{aligned}
e(P)=e(F)
&=\binom{N}{n_1,\ldots,n_m}\prod_{i=1}^m e(F_i)\\
&=\frac{N!}{\prod_i n_i!}
\prod_{i=1}^m\frac{n_i!}{\prod_{v\in P_i}h_F(v)}\\
&=\frac{N!}{\prod_{v\in V}h_F(v)}.
\end{aligned}
\]
\end{proof}

\begin{lemma}[Join multiplicity]
\label{lem:suppl-join-multiplicity}
Every reachable plan $Q$ with $J$ joins satisfies
$|\mathcal L(Q)|\ge 2^J$.
\end{lemma}
\begin{proof}
Fix $\lambda\in\mathcal L(Q)\subseteq\mathcal L(P)$ and a join $j$ with
nonempty input subtrees $Q_L,Q_R$. Let $\lambda_L,\lambda_R$ be the
restrictions of $\lambda$ to their operators. These branches are
incomparable, so every shuffle preserving $\lambda_L$ and $\lambda_R$
represents the same join with unchanged inputs. In particular,
\[
\lambda_L\lambda_R\ne\lambda_R\lambda_L.
\]
Replace the branch subsequence of $\lambda$ by either concatenation,
choosing one different from the original subsequence and keeping all
other positions unchanged. This gives a distinct member of
$\mathcal L(Q)$ representing the same plan.

More generally, writing $a=|Q_L|$, $b=|Q_R|$, and $e(Q)=|\mathcal L(Q)|$,
\[
e(Q_j)=\binom{a+b}{a}e(Q_L)e(Q_R)
\ge 2e(Q_L)e(Q_R).
\]
A unary root leaves $e$ unchanged. Induction over the plan tree therefore
gives one factor of at least two per join, hence $e(Q)\ge 2^J$.
\end{proof}

\begin{proof}[Proof of Proposition~\ref{prop:card}]
First, distinct reachable plans have disjoint linear-extension sets.
Contracting row-wise paths leaves the same fixed skeleton in every plan;
each row-wise operator's admissible positions lie on one path because
join crossings select a unique branch. Thus $Q\ne Q'$ either orders two
operators oppositely within a row-wise path, or places a row-wise operator
on opposite sides of a fixed operator. In either case, some $u,v$ satisfy
\[
u\prec_Q v,\qquad v\prec_{Q'}u,
\]
so $\mathcal L(Q)\cap\mathcal L(Q')=\varnothing$.
Since $\mathcal L(Q)\subseteq\mathcal L(P)$, the preceding lemmas give
\[
\begin{aligned}
2^J|\mathcal R(T)|
&\le\sum_{Q\in\mathcal R(T)}|\mathcal L(Q)|\\
&=\left|\bigcup_{Q\in\mathcal R(T)}\mathcal L(Q)\right|\\
&\le e(P)=\frac{N!}{\prod_{v\in V}h_F(v)}.
\end{aligned}
\]
Dividing by $2^J$ proves
$|\mathcal R(T)|\le\frac{N!}{2^J\prod_{v\in V}h_F(v)}$.
\end{proof}

\subsection{MECE}
\label{app:suppl-mece}

\noindent\textbf{Proposition~\ref{prop:cover}}
Every memoized plan is equivalent to $T$ and memoized once, and
$\mathcal E_0(T)=\mathcal R(T)$.

\noindent\textbf{Memoization} Algorithm~\ref{alg:pd_in} and Algorithm~\ref{alg:pd_across} use set operation $\bigcup$ to iteratively include newly enumerated partitions into a set of seeds while making each member in the set unique. Then it calls the \textsc{Merge} operation to overwrite the current position in the plan $T$ with the new partition(s). 
It is obvious that, when the merged partitions are different, the resulting full plan $T'$ after \textsc{Merge} is also different.
In actual implementation, to prevent duplicates in plan enumeration, we leverage a compact position vector to flag whether the algorithms have generated the same partition before.

\noindent\textbf{Position vector.}
Given an anchor join $j$ and a labeled operator list
$W=(\omega_1,\ldots,\omega_m)$ covering the local transformation.
Number the three arms of $j$ by $0$ (above), $1$ (left), and $2$ (right).
For each $\omega_i$, let $b_i$ be its arm and $r_i\ge1$ its rank counting
outward from $j$. Define
\[
\mathbf p_j(P)=(p_1,\ldots,p_m),\qquad
p_i=3(r_i-1)+b_i+1.
\]
Each entry uniquely determines its arm and rank; hence, for fixed $j,W$, $\mathbf p_j(P)=\mathbf p_j(P')\Leftrightarrow P=P'.$

\begin{lemma}[Coverage]
\label{lem:suppl-partition-coverage}
Starting from the promoted seed, unpruned within- and across-partition
pushdown visits every admissible partition configuration through its
layer-by-layer decomposition, possibly repeatedly.
\end{lemma}
\begin{proof}
For a target $Q$, let $\mathcal V_k$ be the states visited at layer $k$.
Let $\mathcal C_k(Q)$ require agreement with $Q$ on processed chains and
placement of each remaining operator in its highest unprocessed chain
on its path to $Q$. Set
\[
\mathcal F_k(Q)=\mathcal V_k\cap\mathcal C_k(Q).
\]
\begin{enumerate}[label=(\roman*),leftmargin=*,nosep]
\item Assume $Q$ is missing. Since the promoted seed is compatible,
there is a first failure:
\[
\mathcal F_k(Q)\ne\varnothing,\qquad
\mathcal F_{k+1}(Q)=\varnothing.
\]
\item Choose $S\in\mathcal F_k(Q)$. In each current chain $W$, let $D$
contain the operators placed below the join in $Q$ and set $U=W\setminus D$.
Validity of $Q$ implies
\[
x\in W,\ y\in D,\ x\prec_D y\ \Longrightarrow\ x\in D.
\]
\item Thus $D$ is an order ideal. For a topological order $\pi_D$ and
$\pi_U=Q|_U$,
\[
\pi_D\pi_U\in\textsc{ApplyPermute}(dep\_grps).
\]
\item Target validity makes every operator in $D$ crossable, with
producers crossing first. Across-partition pushdown emits the prefix
$|D|$, fixing $U$ and transferring $D$ to its target branches. Cartesian
products of these choices give
\[
S'\in\mathcal F_{k+1}(Q),
\]
contradicting (i). At terminal chains, $D=\varnothing$ and permutation
alone fixes their order. Therefore no admissible $Q$ is missing.
\end{enumerate}
\end{proof}

\begin{proof}[Proof of Proposition~\ref{prop:cover}]
By Lemma~\ref{lem:suppl-partition-coverage}, every partition contributes
all admissible variants. Cartesian products and $\textsc{Merge}$ assemble
every compatible full plan and recursion covers all layers. Since every
transformation is admissible,
\[
\operatorname{supp}\Big({Visited}(T)\Big)=\mathcal R(T).
\]
The complete tuple of labeled partition position vectors uniquely
determines the merged plan. Memoizing only unseen tuples retains exactly
one copy of each visited plan. Hence
\[
\mathcal E_0(T)=\mathcal R(T),
\]
with every member stored once, establishing MECE.
\end{proof}

\section{Speculative Profiling}
\label{app:prof}
We specify the propagation of costs in speculative profiling.

For subquery plan $T$, $C$ is the estimated cost. Operator ${op}$ has input and
output tuples $I_{{op}},R_{{op}}$, result mask $K_{{op}}$,
cost per evaluation $r_{{op}}$, amortization factor $\alpha_{{op}}$,
and uncached evaluation count $n_{{op}}$.
The evaluation-key map $\pi_{{op}}$ uses relevant source-row identities
for filters and projections, full tuple identities for joins or unknown
dependencies.
The cache $ Cache_{{op}}$ maps these keys to filter decisions or
projected values and is empty for uncached operators.
Let $\operatorname{dom}( Cache_{{op}})$ be the cache key set.
For join $j$, $L_j,R_j\ge1$ count outer and inner source
batches, $\operatorname{outer}(j),\operatorname{inner}(j)$ are its child subtrees,
and ${op}_L,{op}_R$ denote the two children of a join.

\begin{algorithm}[t!]
\caption{Cost Propagation}
\label{alg:propagate}
\begin{algorithmic}[1]

\Require $\{K_{{op}},r_{{op}}\}_{{op}\in T}$ from the preceding SLM profiling

\Function{PopulateProfile}{$T,\{K_{{op}},r_{{op}}, Cache_{{op}}\}$}
  \State $C\gets0$
  \State $\alpha_{{op}}\gets1$ for all ${op}\in T$
  \ForAll{joins $j\in T$}
    \State $(L_j,R_j)\gets$ outer and inner source-batch counts, each at least $1$
    \State $\alpha_{{op}}\gets\alpha_{{op}}/R_j$ for ${op}\in{outer}(j)$
    \State $\alpha_{{op}}\gets0$ if the inner side is materialized;
      otherwise $\alpha_{{op}}/(L_jR_j)$, for ${op}\in{inner}(j)$
  \EndFor
  \ForAll{operators ${op}$ in $T$, children before parents}
    \If{${op}$ is a scan}
      \State $R_{{op}}\gets K_{{op}}$
    \Else
      \State $I_{{op}}\gets R_{{op}_L}\times R_{{op}_R}$ for a join;
        otherwise $R_{{child}({op})}$
      \State $R_{{op}}\gets\Call{ApplyMask}{I_{{op}},K_{{op}}, Cache_{{op}}}$
      \If{${op}$ is semantic}
        \State $n_{{op}}\gets|\pi_{{op}}(I_{{op}})
          \setminus\operatorname{dom}( Cache_{{op}})|$

        \State $C\gets C+\alpha_{{op}}\,r_{{op}}\,n_{{op}}$
      \EndIf
    \EndIf
  \EndFor
  \State \Return $C$
\EndFunction
\end{algorithmic}
\end{algorithm}
Algorithm~\ref{alg:propagate} details the \textsc{PropulateProfile} function from line 15 of Algorithm~\ref{alg:saqp}, where the obtained full-batch masks of each operator is used to compute the exact cost of each candidate plan $T$. By applying intersection to the per-operator masks, we can compute costs with any ordering of the set of profiled operators.

\end{document}